%% file: main.tex
\documentclass[letterpaper,twocolumn,10pt]{article}
\usepackage{usenix}

\input{content/environment}

\newcommand{\sysname}{{F\textsc{lare}}}
\newcommand{\groupname}{{Ant Group}}

\begin{document}

\date{}

\title{\Large \bf \sysname{}: Anomaly Diagnostics for Divergent LLM Training in \\ GPU Clusters of Thousand-Plus Scale}

\author{
{\rm Weihao Cui$^{1,2}$\thanks{Equal contribution}, Ji Zhang$^3$\footnotemark[1], Han Zhao$^{1}$\thanks{Corresponding author}, Chao Liu$^3$, Jian Sha$^4$\footnotemark[2],}\\
 {\rm Bingsheng He$^2$,Minyi Guo$^1$, Quan Chen$^1$}\\
 $^1$Shanghai Jiao Tong University, $^2$, National University of Singapore \\ 
 $^3$Independent Researcher, $^4$Ant Group 
} 


\maketitle

\begin{abstract}
The rapid proliferation of large language models has driven the need for efficient GPU training clusters. However, it is challenging due to the frequent occurrence of training anomalies. Since existing diagnostic tools are narrowly tailored to specific issues, there are gaps in their ability to address anomalies spanning the entire training stack. In response, we introduce \sysname{}, a diagnostic framework designed for distributed LLM training at scale. \sysname{} first integrates a lightweight tracing daemon for full-stack and backend-extensible tracing.
Additionally, it features a diagnostic engine that automatically diagnoses anomalies, with a focus on performance regressions.
The deployment of \sysname{} across 6,000 GPUs has demonstrated significant improvements in pinpointing deficiencies in real-world scenarios, with continuous operation for over eight months.
\end{abstract}

\input{content/0-introduction}
\input{content/1-background}

\input{content/2-overview}
\input{content/3-optimizations}
\input{content/4-implementation}

\input{content/5-evaluation}

\input{content/6-related_work}
\input{content/7-conclusion}
\section*{Acknowledgments}
This work is partially sponsored by the National Key Research and Development Program of China (2024YFB4505700),
National Natural Science Foundation of China (62232011) and Natural Science Foundation of Shanghai Municipality (24ZR1430500) and CAAI-Ant Group Research Fund.
We thank the anonymous reviewers and our shepherd Andrew Moore for their constructive feedback and suggestions.

\bibliographystyle{unsrt}
\bibliography{references/others,references/papers}

\end{document}

%% file: content/environment.tex
\usepackage{booktabs}
\usepackage{tikz}
\usepackage{comment}
\usepackage{amsmath}
\usepackage{caption}
\usepackage{subcaption}
\usepackage{graphicx}
\usepackage{float}
\usepackage{algorithm}
\usepackage{algorithmicx}
\usepackage[noend]{algpseudocode}
\algnewcommand{\LineComment}[1]{\State \(//\) #1}
\usepackage[normalem]{ulem}
\usepackage{bbding}
\usepackage{todonotes}
\usepackage{filecontents}
\usepackage{minted}
\usepackage{enumitem}
\usepackage[bottom]{footmisc}
\usepackage[]{hyperref}
\usepackage{cleveref}
\usepackage{pifont}
\usepackage{multirow}
\usepackage[table,xcdraw]{xcolor}
\usepackage{outlines}
\usepackage{diagbox}
\usepackage{wasysym}
\usepackage{titlesec}
\titlespacing*{\section}{0pt}{1ex}{1ex}
\titlespacing*{\subsection}{0pt}{1ex}{1ex}
\titlespacing*{\subsubsection}{0pt}{.1ex}{.1ex}

\definecolor{LightGray}{gray}{0.9}
\definecolor{FadedBanana}{RGB}{255,255,191}
\definecolor{DeepChalk}{RGB}{255,191,191}
\definecolor{FadedFlora}{RGB}{191,255,191}
\definecolor{DeepSnow}{RGB}{191,255,255}
\definecolor{SoapStone}{RGB}{218,218,218}
\definecolor{LightCayenneSixty}{RGB}{239,206,211}
\definecolor{LightCayenne}{RGB}{177,0,28}

\newcommand{\circlenumber}[1]{%
  \tikz[baseline=(char.base)]{%
    \node[shape=circle, fill=black, text=white, draw, 
          inner sep=0.2pt] (char) {\rmfamily #1};%
  }%
}

\makeatletter
\renewcommand{\paragraph}{%
  \@startsection{paragraph}{4}%
  {\z@}{0.ex \@plus 0ex \@minus 0.1ex}{-1em}%
  {\normalfont\normalsize\bfseries}%
}
\makeatother


%% file: content/0-introduction.tex
\section{Introduction}

The advent of large language models (LLMs) has revolutionized the deep learning training community, driving substantial advancements in artificial intelligence-generated content (AIGC). Recognizing their transformative potential to enhance user experiences, leading corporations are proactively leveraging LLMs to enhance a wide array of user-oriented services~\cite{antgroup, jiangMegaScaleScaling,dongEnhancingLargescale}.
To meet the computational demands of LLM training, they construct large-scale training clusters comprising the latest GPUs interconnected via high-bandwidth links.

\begin{figure}
    \centering
    \includegraphics[width=.95\linewidth]{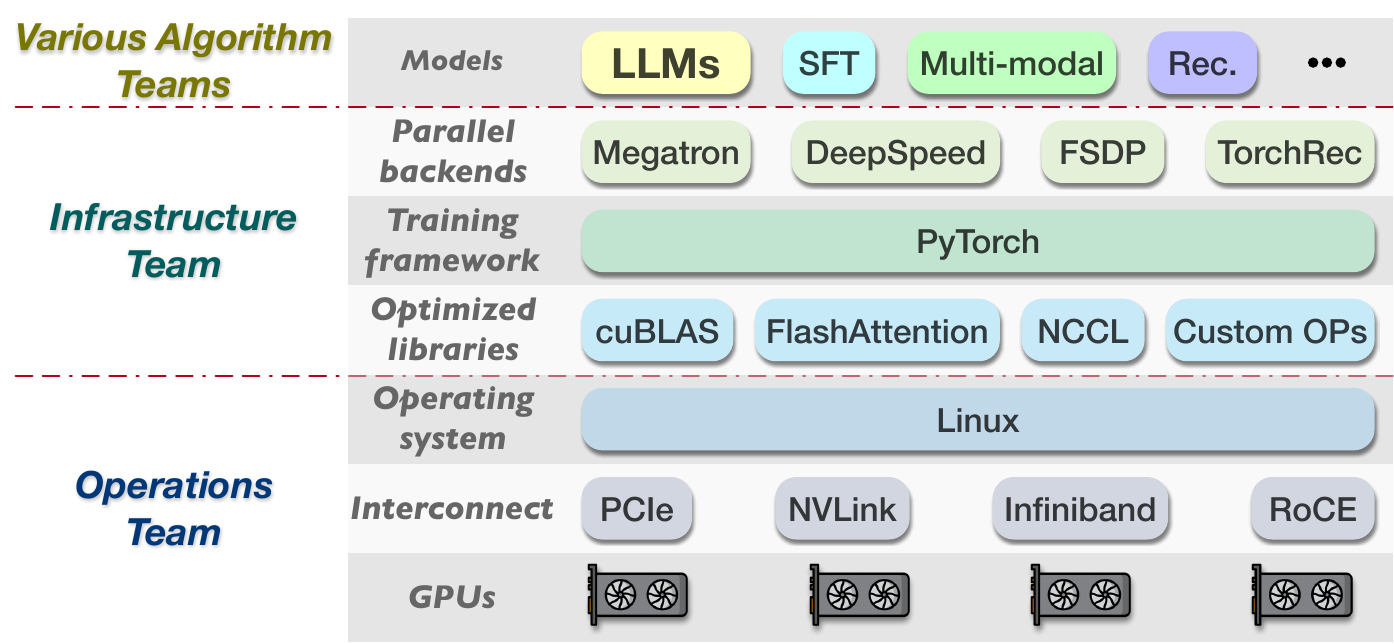}
        \vspace{-2mm}
    \caption{The training stack of large-scale training cluster.}
    \label{fig:training-stack}
    \vspace{-5mm}
\end{figure}

\autoref{fig:training-stack} depicts the general training stack of the large-scale training cluster in modern corporations.
As shown, the operations team manages low-level resources, and the infrastructure team delivers training optimizations\cite{cublas2024,nccl2024,daoFlashAttentionFast,paszkePyTorchImperative}, with particular emphasis on parallel backends\cite{zhaoPyTorchFSDP,shoeybiMegatronLMTraining,rajbhandariZeROMemory}.
Supported by these two teams, \textbf{various algorithm teams} focus on adapting LLMs for user-facing applications through various training methods~\cite{shengHybridFlowFlexible,hanParameterEfficientFineTuning}.
Notably, the training cluster also supports other deep learning jobs, such as recommendation models and their specific parallel backend, TorchRec~\cite{ivchenkoTorchRecPyTorch}.

Efficient large-scale distributed LLM training requires effective coordination of software and hardware across the stack.
However, training anomalies frequently arise from various layers.
These anomalies include obvious job failures and non-obvious training slowdowns, which can be categorized into three types: performance regressions, job failures, and fail-slows. 
Performance regressions may result from incorrect training settings (addressed by algorithm teams) or unoptimized operators (addressed by infrastructure teams). In contrast, most job failures and fail-slows arise from hardware issues, which are handled by the operations team. 
\textbf{One key insight in this paper is that fail-slows are sudden and easy-to-detect slowdowns due to transient issues in components, whereas performance regressions are persistent and hard-to-detect slowdowns caused by code updates or configuration drift.
}

Drawing on our daily operational experiences, teams have spent significant time in cross-team collaborations to address these anomalies, often repeating efforts on similar cases.
For instance, performance regressions caused by unnecessary synchronizations introduced by algorithm teams often require collaboration with the infrastructure team.
Yet, eliminating these synchronizations usually only involves deleting a few lines of code in the training scripts. Therefore, a dedicated diagnostic framework is urgently needed to mitigate this.

The framework first should collect sufficient runtime data for anomaly detection and potential causes identification, enabling teams to resolve them with minimal collaboration.
To this end, runtime data must be continuously gathered from all layers of the training stack, especially from the software in the algorithm and infrastructure teams.
Secondly, root causes should be narrowed as much as possible; otherwise, even if recurrent anomalies are routed to the right teams, they may still be unable to resolve them independently.

To design such a diagnostic framework, we identify two main challenges.
\textit{\textbf{C-1}: Designing a full-stack tracing mechanism with backend extensibility is challenging.}
The tracing mechanism collects runtime data for anomaly diagnosis.
While full-stack tracing is needed to capture sufficient runtime data, diverse parallel backends calls for backend-extensibility, 
However, these requirements often conflict, as full-stack tracing usually intrudes into parallel backend codebases, reducing backend-extensibility.
\textit{\textbf{C-2}: Detecting and diagnosing root causes is challenging.}
Errors or fail-slows in LLM training often show similar symptoms, such as process hangs or sudden drops in training speed.
Worse, regressions cannot be directly confirmed by comparing macro metrics, such as current training throughput, with historical runs, since throughput declines may be expected.
These factors make it difficult to detect anomalies and narrow down their root causes.

Faced with these challenges, previous efforts\cite{wuGREYHOUNDHunting,jiangMegaScaleScaling,dongEnhancingLargescale} targeting distributed LLM training fail to provide comprehensive solutions.
This is because they are narrowly designed for specific problems or scenarios, making them inadequate for addressing issues across the software-hardware stack.
For example, Greyhound~\cite{wuGREYHOUNDHunting}, C4D~\cite{dongEnhancingLargescale}, and Holmes~\cite{yaoHolmesLocalizing} can only trace and diagnose errors or fail-slows under the responsibility of operations team.
{\bf However, they could not solve the performance regressions originated from upper-level teams.}
Besides, tools such as MegaScale~\cite{jiangMegaScaleScaling} are optimized for pre-training scenarios involving a single algorithm team and a single backend.
It intrudes into the backend codebase to achieve full-stack tracing, but {\bf  this tight coupling makes it difficult to plug into other parallel backends.}
Moreover, MegaScale only provides visualization of distributed training and relies on cross-team collaboration and manual effort to pinpoint root causes for detected anomalies.

To this end, we present \sysname{}, an anomaly diagnostic framework for divergent LLM training in GPU clusters at thousand-plus scale.
It places particular emphasis on diagnosing performance regression anomalies from algorithm and infrastructure teams.
\sysname{} consists of two components: a per-training-process tracing daemon and a diagnostic engine.
To collect sufficient runtime data with backend-extensibility (\textbf{C-1}), the tracing daemon selectively instruments key code segments in a plug-and-play manner across both Python and C++ runtimes.
Most importantly, it leverages CPython tracing mechanisms to trace Python code segments without intruding into backbone codebases.
It also provides an easy-to-play interface—simply adding shell environments—for extending to new backends and allowing all teams to selectively trace Python code segments.

With real-time data collected by the tracing daemon, the diagnostic engine detects and diagnoses anomalies while narrowing their root causes (\textbf{C-2}).
As for error diagnostics, \sysname{} introduces a novel intra-kernel inspecting mechanism, providing fine-grained diagnostics specifically targeting communication-related hang errors.
It enables $\mathcal{O}(1)$-complexity faulty machine diagnostics without exhaustive or blind searches based on NCCL tests.
As for slowdown diagnostics, \sysname{} proposes holistic aggregated metrics that encompass not only commonly used macro metrics like training throughput but also novel micro metrics, such as issue latency distribution.
Importantly, the new proposed micro metrics enable \sysname{} to detect and diagnose regressions.
Overall, with these metrics, \sysname{} enables teams to independently address those routed slowdowns, including subtle ones such as a $2.66\%$ regressions in real-world workloads.

We extensively evaluated \sysname{} in terms of its runtime overhead. \sysname{} incurs an average latency overhead of only $0.43\%$ across various LLMs and backends on 1024 H800 GPUs. Meantime, \sysname{} only generates just $1.5\text{MB}$ of tracing logs per GPU in a real-world model trained on 1536 H800 GPUs.
\sysname{} has been deployed in our training cluster at Ant Group\cite{antgroup}, utilizing over 6,000 GPUs to date.
The greatest benefit of \sysname{}'s deployment is that it eases the handling of recurrent anomalies.
\sysname{} has been open-sourced at \href{https://github.com/intelligent-machine-learning/dlrover/tree/master/xpu_timer}{DLRover}.
It serves as a core component of DLRover~\cite{dlrover}, an automated distributed DL system deployed within \groupname{} and supported by the LF AI \& Data Foundation~\cite{lfaidata}. Our contributions are as follows.
\begin{itemize}
[leftmargin=*,topsep=0.1em,itemsep=-0.5em]
    \item We highlight the urgent need for a real-time, holistic diagnostic framework capable of identifying LLM training anomalies across the entire training stack.
    \item We introduce \sysname{}, a diagnostic framework specifically designed to tackle the critical challenges of full-stack tracing, root cause diagnosis, and backend extensibility in LLM training diagnostics.
    \item We deploy \sysname{} across more than 6,000 GPUs over an 8-month period, deriving typical case studies and practical insights from its daily operations.
\end{itemize}

%% file: content/1-background.tex
\begin{table*}
\centering
\scriptsize
\caption{A comprehensive analysis of anomalies encountered in \groupname{}, with \sysname{}’s primary target highlighted in yellow.}
\vspace{-1mm}
\label{tb:anomaly-analysis}
\begin{tabular}{c|ccccccccc}
\hline\hline
\diagbox{}{} & \multicolumn{9}{c}{\textbf{Anomalies}}                                                                                                                                                                                                                                                                                                                                                                                                                                                                                                                                                                                                                                                                                                                                                                                                                                                                                                                                                                                                                                                                                                                                                              \\ \hline
\textbf{Type}                   & \multicolumn{3}{c|}{Error}                                                                                                                                                                                                                                                                                                                                                                                                                                        & \multicolumn{6}{c}{Slowdown}                                                                                                                                                                                                                                                                                                                                                                                                                                                                                                                                                                                                                                                                                                     \\ \hline
\textbf{Taxonomy}      & \multicolumn{1}{c|}{\begin{tabular}[c]{@{}c@{}}OS\\ errors\end{tabular}} & \multicolumn{1}{c|}{\begin{tabular}[c]{@{}c@{}}GPU\\ errors\end{tabular}} & \multicolumn{1}{c|}{\cellcolor[HTML]{FFFFC7}{\color[HTML]{333333} \textbf{\begin{tabular}[c]{@{}c@{}}Network\\ errors\end{tabular}}}} & \multicolumn{1}{c|}{\cellcolor[HTML]{FFFFC7}{\color[HTML]{333333} \textbf{\begin{tabular}[c]{@{}c@{}}New\\ algorithms\end{tabular}}}} & \multicolumn{1}{c|}{\cellcolor[HTML]{FFFFC7}{\color[HTML]{333333} \textbf{\begin{tabular}[c]{@{}c@{}}Unnecessary\\ synchronization\end{tabular}}}} & \multicolumn{1}{c|}{\cellcolor[HTML]{FFFFC7}{\color[HTML]{333333} \textbf{\begin{tabular}[c]{@{}c@{}}Un-optimized\\ kernels\end{tabular}}}} & \multicolumn{1}{c|}{\cellcolor[HTML]{FFFFC7}{\color[HTML]{333333} \textbf{\begin{tabular}[c]{@{}c@{}}Memory\\ management\end{tabular}}}} & \multicolumn{1}{c|}{\begin{tabular}[c]{@{}c@{}}GPU\\ underlocking\end{tabular}} & \begin{tabular}[c]{@{}c@{}}Network\\ jitter\end{tabular} \\ \hline
\textbf{Symptom}          & \multicolumn{3}{c|}{\begin{tabular}[c]{@{}c@{}}Runtime hang\\ or crash error\end{tabular}}                                                                                                                                                                                                   & \multicolumn{4}{c|}{\cellcolor[HTML]{FFFFC7}{\color[HTML]{333333} \textbf{\begin{tabular}[c]{@{}c@{}}Regressions compared to\\ historical jobs\end{tabular}}}}                                                                                                                                                                                                                                                                                                                                                                                            & \multicolumn{2}{c}{\begin{tabular}[c]{@{}c@{}}Fail-slows compared to\\ prior training steps\end{tabular}}        \\ \hline
\textbf{Team}             & \multicolumn{3}{c|}{Operations}                                                                                                                                                                                                                                                              & \multicolumn{2}{c|}{\cellcolor[HTML]{FFFFC7}{\color[HTML]{333333} \textbf{Algorithm}}}                                                                                                                                                                                                     & \multicolumn{2}{c|}{\cellcolor[HTML]{FFFFC7}{\color[HTML]{333333} \textbf{Intrastructure}}}                                                                                                                                                                                            & \multicolumn{2}{c}{Operations}
  \\ \hline\hline
\end{tabular}
\vspace{-4mm}
\end{table*}

\section{Background and Motivation}

\subsection{Large-Scale LLM Training Stack}
Referring to the software-hardware stack 
in \autoref{fig:training-stack}, we delve into how LLMs are reshaping teams in leading corporations.

\paragraph{Advancing of LLM applications.}
With the advancing of LLM algorithms~\cite{deepseekmoe}, LLMs excel not just in natural language understanding, but also in tackling advanced tasks such as multimodal tasks ~\cite{vit,sora}, reasoning tasks~\cite{openaio1,rstar}. For instance, customer service teams fine-tune LLMs to develop chatbots that generate accurate responses to complicated questions. Similarly, product development teams leverage multimodal LLMs to generate comprehensive product descriptions by integrating inputs, such as product images and textual data. 
Consequently, different algorithm teams continue to innovate LLM models tailored for diverse application scenarios.

\paragraph{Advancing of training cluster.}
Training these LLMs is computationally intensive, requiring large-scale GPU clusters of thousand-plus scale~\cite{llama3-1}.
Therefore, leading corporations are continuously investing in large training clusters powered by state-of-the-art hardware~\cite{xai_colossus_training_cluster,meta_ai_training_infrastructure}.
These clusters not only expand in scale but also incorporate the latest GPUs~\cite{nvidia_ampere_architecture,nvidia_hopper_architecture} with higher computational performance and advanced interconnect~\cite{nvidia_nvswitch,nvidia_nvl72}.
Efficient operation of these large-scale clusters is critical to the corporations, which requires the operations team to ensure uninterrupted training performance.

\paragraph{Advancing of training infrastructure.}
To enable easy, efficient, and scalable LLM training within large-scale clusters, the infrastructure teams build the dedicated software stack. This stack bridges the gap between algorithm teams and operations team. Specifically, the infrastructure team focuses on optimizing the software stack by integrating advanced operator libraries~\cite{cublas2024,cutlass2024,nccl2024}, training framework~\cite{paszkePyTorchImperative}, and state-of-the-art model parallel backends~\cite{zhaoPyTorchFSDP,shoeybiMegatronLMTraining,liangTorchTitanOnestop,ivchenkoTorchRecPyTorch}.
Parallel backends are fundamental for efficiently training LLMs, enabling tensor, data, and pipeline parallelism across thousands of GPUs.

\subsection{Anomalies of Large-scale LLM Training}
\label{sec:anomalies}
Due to the complexity of the entire training stack, various issues can easily occur.
These issues affect both the training speed of individual training jobs and the overall utilization of training clusters, collectively termed as anomalies.

\autoref{tb:anomaly-analysis} presents a distilled analysis of the common anomalies in our real-world cluster, broadly categorized into two primary types: runtime errors and slowdowns.
Based on symptom differences, slowdowns can be further divided into two types: performance regressions (persistent slowdowns) and fail-slows (sudden and acute slowdowns).

Based on three months of traces from a cluster with 6000+ GPUs, there are 127 errors and 135 slowdowns observed across 3047 jobs. The slowdowns include 78 performance regressions and 57 fail-slows. When such anomalies occur, they are first reported by algorithm teams, and then multiple teams coordinate and spend significant effort to resolve them.


Specifically, for runtime errors and fail-slows, the operations team intervenes to resolve these acute and easy-to-detect job failures. Without timely intervention, training jobs may fail or run at very low throughput, significantly reducing training efficiency.
For performance regressions, algorithm teams are often unfamiliar with why these regressions occur with their own code changes, so collaboration with infrastructure teams is required to resolve them. In a training stack like \autoref{fig:training-stack}, the situation worsens: the infrastructure team supports multiple algorithm teams, each potentially using different parallel backends for LLM training. As recurrent regressions arise across different algorithm teams, the infrastructure team must repeatedly expend effort to address them.

This dilemma motivates a dedicated diagnostic framework to streamline anomaly resolution, especially performance regressions. To achieve this, we must collect sufficient runtime data across the training stack. For example, overall training throughput helps detect fail-slows, while per-operator performance reveals regressions introduced by minor software changes. We must also seamlessly support multiple parallel backends used by different algorithm teams. Finally, for detected anomalies, we aim to narrow down their root causes as much as possible, enabling them to be routed to the appropriate team and resolved independently without unnecessary cross-team collaboration.

However, designing such a framework is far from trivial, posing two key challenges.

\paragraph{The difficulty of full-stack, backend-extensible tracing.}
Diagnosing performance regressions across diverse parallel backends requires full-stack and backend-extensible tracing to collect sufficient runtime data.
These goals often conflict. 
Full-stack tracing tends to be intrusive to each backend’s codebase, whereas extensibility demands minimal changes to the backends.
A typical example is MegaScale~\cite{jiangMegaScaleScaling}, which enables low-overhead tracing for FSDP--the native PyTorch backend--by patching the FSDP codebase.
Extending support to other backends would require similar patches.
A tracing design for runtime data collection in training jobs of multiple algorithm teams should avoid such backend-intrusive modifications.

\paragraph{The difficulty of narrowing down attribution.}
Errors and fail-slows in \autoref{tb:anomaly-analysis}  often show obvious symptoms, as the job either fail or show sudden training speed degradation.
There are various works like C4D~\cite{dongEnhancingLargescale}, Greyhound~\cite{wuGREYHOUNDHunting}, and Holmes~\cite{yaoHolmesLocalizing} that are proposed to diagnose them.
However, automatic performance regressions are rarely explored in prior work.
Attributing them is more challenging than diagnosing fail-slows.
First, they are harder to detect. Regressions are persistent slowdowns caused by software changes, and monitoring training throughput alone is insufficient. Algorithm teams may not notice them and may assume jobs are running normally.
Second, linking a detected regression to the corresponding software change is difficult. For example, whether Python garbage collection causes a regression depends on how it is invoked.
Thus, new metrics must be designed based on operational expertise, especially from infrastructure team, to detect and diagnose regressions.

%% file: content/2-overview.tex
\section{\sysname{} Design}
\begin{figure}
    \centering
    \includegraphics[width=\linewidth]{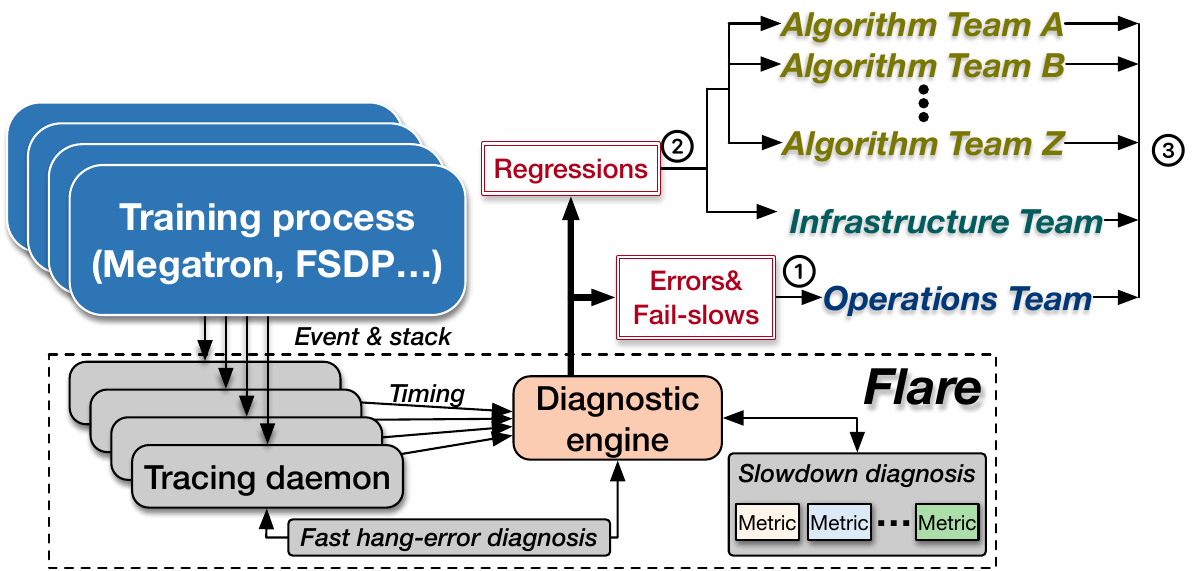}
    \vspace{-1mm}
    \caption{Architecture overview of \sysname{}.}
    \label{fig:overview}
    \vspace{-4mm}
\end{figure}



In this paper, we propose \sysname{}, a diagnostic framework for diverse large-scale LLM training.
It places particular emphasis on diagnosing performance regression automatically from algorithm and infrastructure teams.
\autoref{fig:overview} illustrates the architecture of \sysname{}, deployed in Ant Group’s large-scale training cluster. \sysname{} is composed of two components: the tracing daemon and the diagnostic engine. 

By automatically attaching a tracing daemon to each training process in LLM jobs, \sysname{} provides a lightweight tracing mechanism.
The daemon is backend-extensible but also enables the collection of sufficient runtime data across the entire training stack.
For each job, runtime data such as function timing events and call stacks are intercepted for monitoring and diagnostics.
The diagnostic engine employs a fast hang-error diagnostic method and leverages novel aggregated metrics to effectively identify slowdowns, especially regressions.
The aggregated metrics encompass both macro-level metrics, such as training throughput, and micro-level metrics, including issue latency distribution of GPU kernels.

The overall diagnostic pipeline is as follows:

\ding{172} Fast hang-error diagnosis enables error detection and analysis. Fail-slows are detected through macro metrics and validated by micro metrics. Errors and fail-slows are then routed to the operations team for independent handling.

\ding{173} Regressions are detected by monitoring the proposed micro metrics and diagnosed via Python API analysis. \sysname{} routes potential regressions with narrowed root causes to the algorithm or infrastructure team.

\ding{174} The algorithm, infrastructure, and operations teams collaborate only when anomalies cannot be resolved independently by the routed team.



%% file: content/3-optimizations.tex
\section{Lightweight Selective Tracing}
The runtime data collection overhead primarily stems from high memory usage rather than interference with computing resources, since profiling APIs like CUPTI\cite{cupti2024} can operate in a background thread.
For instance, profiling a Llama-70B model trained on 512 H800 GPUs using PyTorch’s built-in profiler produces a log file of $5.5\text{GB}$ (in JSON format, compressed to $451\text{MB}$) for each training step.
This substantial memory overhead renders such arbitrary profiling methods impractical for continuously collecting real-time data.

Therefore, \sysname{} selectively instruments code segments of key APIs and kernels to collect real-time information. 
This design is based on an insight into LLM training on large-scale GPUs: LLM training is predominantly dominated by a limited set of deep learning operators.
These operators mainly include matrix multiplication and cross-GPU communication operators. 
In the rest of this section, we introduce the way \sysname{} achieves full-stack and backend-extensible tracing through employing two techniques: plug-and-play instrumentation and timing with stack reconstruction.


\subsection{Plug-and-Play Instrumentation}
\label{sec:tracing}


\begin{figure}
    \centering
    \includegraphics[width=.78\linewidth]{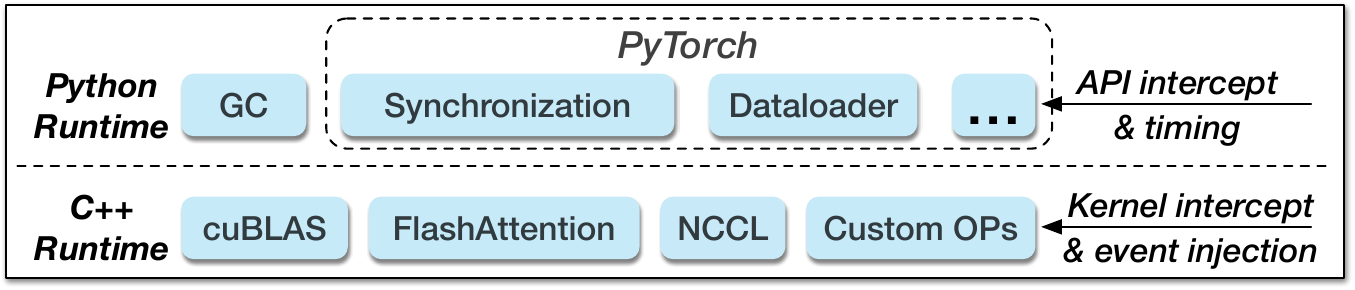}
        \vspace{-1mm}
    \caption{Instrumented key code segments in \sysname{}.}
    \label{fig:key-segment}
    \vspace{-4mm}
\end{figure}

Full-stack, backend-extensible tracing requires plug-and-play instrumentation that captures the required runtime data without intrusive backend changes.
\autoref{fig:key-segment} shows the code segments instrumented by \sysname{}, which are widely used in backend codebases and broadly fall into two groups.
The first intercepts key API calls, including Python’s garbage collection (GC), PyTorch’s dataloader, and GPU synchronization.
The second targets critical GPU computation and communication kernels executed at the C++ runtime level. These kernels dominate the workload in large-scale training.

Intercepting C++ functions is straightforward with backend extensibility, as mechanisms like \texttt{LD\_PRELOAD} can hook any dynamically linked C++ functions.
In contrast, intercepting Python APIs is more challenging when enforcing backend extensibility. A common method for tracing is to provide interfaces such as Python decorators to wrap APIs in backend.
E.g., MegaScale supports FSDP by patching PyTorch, which requires intruding into each backend for full-stack support.

To avoid such intrusiveness for backend extensibility, \sysname{} adopts a CPython-based intercepting mechanism. It maintains a list of tracing-required APIs for each backend.
When a training job starts, \sysname{} imports these APIs and retrieves their bytecode.
During long training runs, it intercepts them directly using CPython’s profiling API \texttt{PyEval\_SetProfile} based on the bytecode.

The above design allows \sysname{} to instrument either Python or C++ functions in a plug-and-play manner.
For easy plug-in, \sysname{} does not modify any backend codebase before tracing.
For easy play, \sysname{} exposes user-friendly interfaces for tracing new functions.
Specifically, users only need to configure environment variables before launching jobs, such as \texttt{export TRACED\_PYTHON\_API=``torch.cuda@synchronize''}.

Intercepting C++ functions requires explicit registration through a C++ interface.
This is practical because the infrastructure team develops both these C++ functions and \sysname{}, ensuring seamless integration and functionality.

\subsection{Timing with Stack Reconstruction}
\begin{figure}
    \centering
    \includegraphics[width=0.8\linewidth]{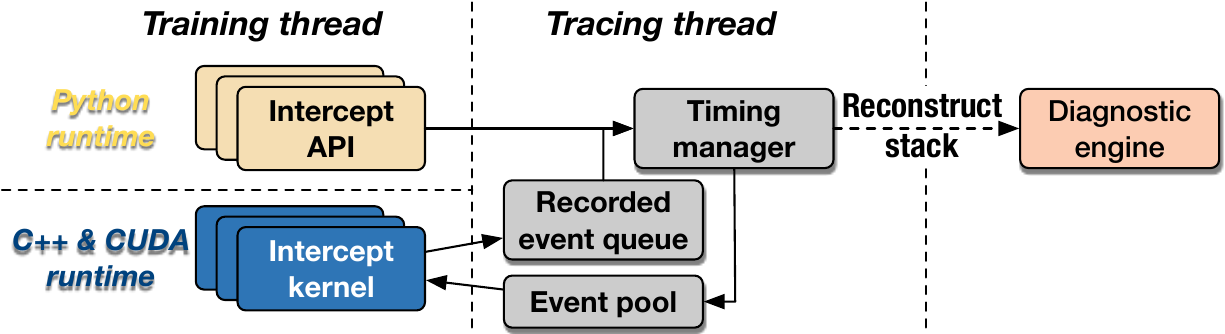}
        \vspace{-3mm}
    \caption{Timing the training in the background.}
    \label{fig:timing}
    \vspace{-5mm}
\end{figure}

With intercepted Python APIs and GPU kernels, FLARE measures their elapsed latencies, as shown in \autoref{fig:timing}. Specifically, a dedicated tracing thread runs in the background to efficiently manage timing data. It employs different timing mechanisms for Python APIs and GPU kernels.

For synchronous Python API calls, FLARE directly records their start and end timestamps and forwards them to the timing manager.
For GPU kernels, which execute asynchronously, FLARE injects CUDA events\cite{cudaevents2024} after an interception to record execution status. These events are enqueued for further processing. The timing manager queries the status of the queued events in the background, avoiding any disruption to the training thread. Additionally, during GPU kernel interception, FLARE extracts input specifications, such as memory layout, to support subsequent anomaly diagnostics.


As training progresses, the timing manager proactively streams real-time data to \sysname{}’s diagnostic engine. However, since plug-and-play instrumentation uses separate methods for Python APIs and C++ functions, it omits the call stack linking them, which is essential for anomaly diagnostics. Fortunately, we record not only their durations but also their start and end timestamps. Using this information, the tracing thread reconstructs the call-stack relationships before sending the data to the diagnostic engine.

\section{Anomaly Detection and Diagnosis}

In this section, we present \sysname{}’s automated diagnosis of anomalies: runtime errors, fail-slows, and performance regressions. Among these, \sysname{} pays particular attention to regressions, which have not been explored in prior work.

\subsection{Fast Runtime Error Diagnosis}
\label{sec:diagnose_error}

Runtime errors often stem from issues like operating system crashes, GPU failures, or network disruptions, which can generally be resolved by isolating the problematic machines and restarting the training job.
A typical symptom associated with these errors is the hanging of the training job.
Training LLMs across numerous GPUs in a distributed manner inherently relies on the coordination of training processes.
When the aforementioned errors occur, they rarely affect all training processes simultaneously.
The training process hangs if a GPU fails.
In this context, \sysname{} focuses on rapidly diagnosing hang errors by identifying faulty machines.
Then, \sysname{} routes this information to the operations team, enabling the training job to restart with healthy machines.

Specifically, the diagnostic engine detects hang errors by examining the status of tracing daemons.
The daemon operates in the background of the training thread and continuously queries events recorded during job execution.
If it fails to confirm the completion of an event within a predefined timeout interval, it proactively reports a potential hang error to the diagnostic engine.
If a tracing daemon transmits none real-time data within the specified timeout interval, the diagnostic engine also interprets this as an indication of a hang error.

After hang errors are reported, they are classified as either communication or non-communication errors. \sysname{} diagnoses these errors in two steps: first, a coarse-grained diagnosis through call stack analysis; and second, a fine-grained diagnosis using intra-kernel inspecting.

\begin{figure}
    \centering
    \includegraphics[width=.84\linewidth]{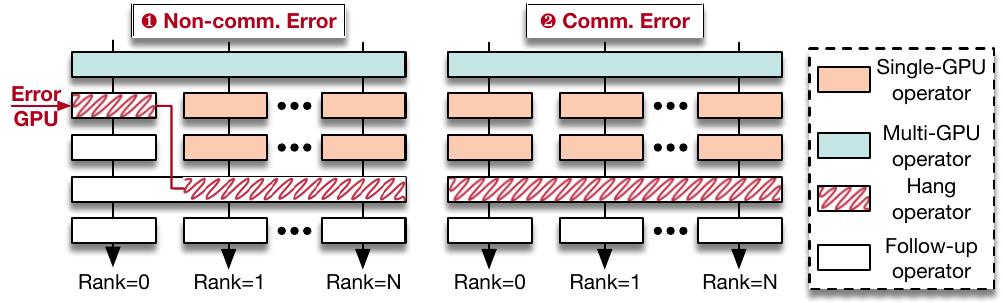}
        \vspace{-3mm}
    \caption{Diagnosing hang errors via call stack analysis.}
    \label{fig:hang-non-comm}
    \vspace{-5mm}
\end{figure}

\paragraph{Diagnosis using call stack analysis.} 
This diagnosis is used to identify problematic machines encountering non-communication errors. 
\autoref{fig:hang-non-comm} illustrates an example of hang-error diagnosis via call stack analysis.
As shown in the left of \autoref{fig:hang-non-comm}, when the training process of rank-$0$ crashes or is suspended due to a non-communication error, it halts at a call stack corresponding to a non-communication function.
In contrast, the training processes of other ranks continue executing correctly and eventually stop at a call stack associated with a communication-related function that depends on coordination with rank-$0$.
In this scenario, the machine associated with rank-$0$ is identified as the source of the error.

However, communication hang cannot be identified through call stack analysis.
As shown in the right of \autoref{fig:hang-non-comm}, the training processes of all ranks terminate at the same call stack of a communication function, such as allreduce or allgather.
We further investigate the symptoms of communication hang errors and obtain two observations. Firstly, some communication hang errors generate error logs. For instance, if the link between RDMA NICs breaks, an error code of \texttt{12} is produced. Secondly, more hang errors result in an endless loop within the launched communication kernels, ultimately leading to job termination after a predefined timeout.

To identify the faulty machine responsible for such errors, a common approach is to terminate the training process and run NCCL tests across all GPUs.
However, in large-scale training where multiple model parallelisms--such as tensor~\cite{shoeybiMegatronLMTraining}, pipeline~\cite{narayananPipeDreamGeneralized}, and expert parallelism~\cite{lepikhinGshardScaling}--are combined, the NCCL tests must span all configured communication groups.
In our experience, this exhaustive and blind search can take over half an hour at the thousand-GPU scale.

\begin{figure}
    \centering
    \includegraphics[width=.9\linewidth]{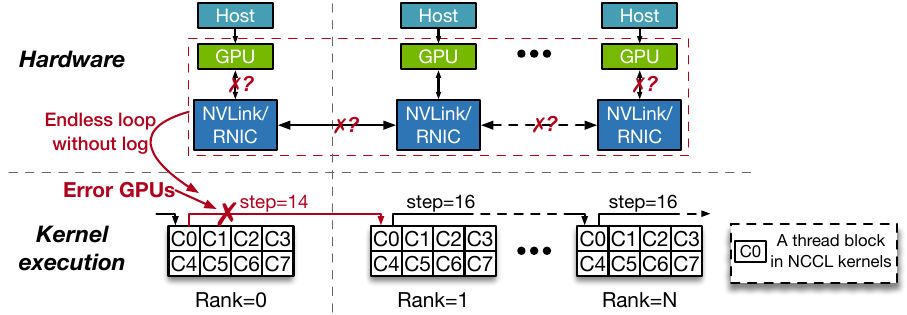}
        \vspace{-3mm}
    \caption{Diagnosing hang errors via intra-kernel inspecting.}
    \label{fig:hang-comm}
    \vspace{-5mm}
\end{figure}

\begin{figure*}
    \centering
    \includegraphics[width=.81\linewidth]{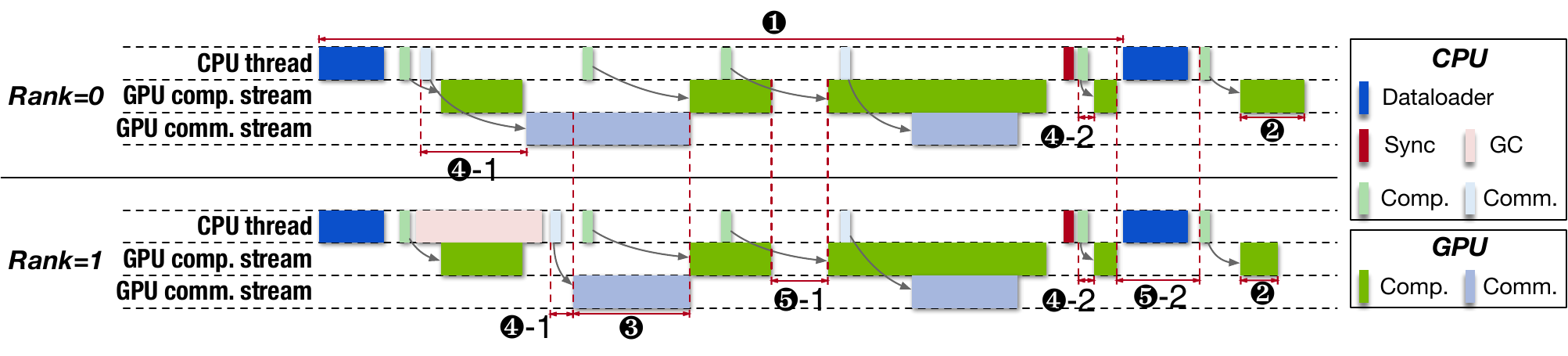}
    \vspace{-3mm}
    \caption{A timeline of a distributed training job annotated with aggregated metrics used for diagnosing slowdowns in \sysname{}.}
    \label{fig:aggregated-metric}
    \vspace{-5mm}
\end{figure*}

\paragraph{Diagnosis using intra-kernel inspecting.}
Faced with this problem, \sysname{} introduces a minute-level diagnostic approach using intra-kernel inspecting with $\mathcal{O}(1)$ complexity.
Instead of stopping the hanged training processes and blindly probing with NCCL tests, \sysname{} fully utilizes real-time data at the error point for diagnosis.
Specifically, the intra-kernel inspecting leverages CUDA-GDB, the debugging tool for CUDA programming.
It instructs the tracing daemon to attach to the halted training processes using CUDA-GDB.
Once attached, the tracing daemon executes a script capable of automatically extracting detailed communication statuses to identify unhealthy machines.

\autoref{fig:hang-comm} depicts an example of diagnosing communication hang errors in a hanging ring-allreduce kernel.
In the ring-allreduce kernel, each thread block is responsible for transmitting data between linked adjacent ranks within the kernel’s constructed ring.
The data are split into chunks and thread blocks of adjacent ranks work together to transmit the chunks step by step.
Thus, \sysname{} retrieves the register values corresponding to the loop steps used for data transmission between linked ranks. 
Theoretically, the connection with the minimum step reveals the related GPUs experiencing errors.
This intra-kernel inspecting process is performed in parallel across all involved GPUs.
As a result, its complexity is $O(1)$, enabling completion within a few minutes.

\sysname{} then routes the diagnostic information for detected errors to the operations team, assisting with tasks such as isolating faulty machines and restarting the training job.
Notably, this mechanism does not degrade the performance of the training job.
The tracing daemon directly inspects the SASS code to retrieve register values.
Thus, \sysname{} does not require recompiling the NCCL library with debugging information enabled, which would disables several compiler optimizations.
Moreover, \sysname{} attaches to the training processes only after the job hangs, without interfering with normal execution.

\subsection{Aggregation for Slowdown Diagnosis}
\label{sec:slowdown-anomalies}

To holistically identify fail-slow and regression slowdowns, \sysname{} aggregates real-time data from the tracing daemon into five primary metrics, shown in \autoref{fig:aggregated-metric}.
\textit{These metrics reflect the consensus that a ``health'' training pipeline should exhibit a timeline saturated with GPU kernels dedicated to computation or communication.}
Computation kernels should deliver high FLOPS, while communication kernels should sustain high bandwidth. Deviations from these characteristics indicate idle GPU resources, signaling potential slowdowns in training jobs.
In the rest of this section, we introduce the five metrics and explain how they are used to detect and narrow down the root causes of fail-slows and regressions.


\subsubsection{Detecting Fail-slows with Macro Metric}
\paragraph{\protect\circlenumber{1} Training throughput for detecting fail-slows.}
Fail-slows caused by low-level hardware changes, such as GPU underclocking or network jitter, are apparent and can be detected solely through comparisons across training steps.
To this end, \sysname{} measures training throughput by timing the rate at which input data is consumed by the training pipeline.
It is achieved by instrumenting the dataloader API of Pytorch.

\subsubsection{Detecting Regression with Micro Metrics}
Regressions caused by software changes introduced by algorithm and infrastructure teams are often subtle and difficult to detect. Identifying them usually requires manual comparisons across historical training jobs. However, directly comparing training throughput between monitored and historical jobs does not reliably reveal regressions, since model size, input datasets, and other factors may differ significantly. To address this, \sysname{} detects regressions using micro metrics.
This is because when macro metrics do not show sudden changes, abnormalities in micro metrics often indicate potential persistent performance regressions.

\paragraph{\protect\circlenumber{2} FLOPS and \protect\circlenumber{3} bandwidth.}
FLOPS and bandwidth capture the performance of instrumented computation and communication kernels, and can be used to detect regressions.
For example, computation kernels with large input sizes but much lower FLOPS than the theoretical value indicate a regression in computation performance.

\sysname{} monitors the FLOPS of instrumented critical computation kernels, leveraging timing data and input layout.
It also monitors the bandwidth of communication kernels.
A communication operator requires launching the communication kernels on all ranks. Since variations in kernel-issue timestamps exist across different ranks, \sysname{} calculates the communication bandwidth by utilizing the start and end timestamps of the final communication kernels issued across all participating ranks.

Notably, when analyzing FLOPS and bandwidth to detect regressions, \sysname{} accounts for the overlap of communication and computation kernels, which is common in training MoE-based LLMs~\cite{deepseek-aiDeepSeekV3Technical}.
This ensures that computation kernels with falsely low FLOPS are not mistakenly flagged.

While FLOPS and bandwidth ensure that both critical computation and communication GPU kernels operate at high performance, they do not cover the less critical operations, such as various CPU operations and element-wise activation GPU kernels. 
Meantime, \sysname{}’s tracing also omits the monitoring of these operations. 
To detect their potential contributions to regressions, we further classify these not-instrumented operations into three categories: intra-step CPU operations, inter-step CPU operations, and minority GPU kernels.
Intra-step CPU operations and inter-step CPU operations differ due to their occurrences within the timeline of training steps.
Minority GPU kernels refer to those GPU kernels that often occupy little GPU computation resources.

Specifically, two metrics are introduced for regression detection: issue latency distribution for intra-step CPU operations and void percentage for inter-step CPU operations and minority GPU kernels.

\paragraph{\protect\circlenumber{4} Issue latency distribution for kernel-issue stall.}\label{sec:diagnose:issue}
In a well-optimized parallel backend, only the necessary intra-step CPU operations for launching GPU kernels or coordinating the training processes are expected. However, algorithm teams may inadvertently introduce unnecessary GPU synchronizations when modifying the LLM model. Meantime, certain function calls, such as GC~\cite{jiangMegaScaleScaling,shoeybiMegatronLMTraining}, may be implicitly triggered by the Python runtime. These intra-step CPU operations can occur repeatedly during the model’s forward pass, bringing considerable overhead. In such cases, these operations cause a regression anomaly known as a kernel-issue stall, leading to GPU idle time within the training step.

\protect\circlenumber{4}--1 in \autoref{fig:aggregated-metric} shows the example of Python runtime GC.
In the figure, the Python runtime GC stalls the CPU thread and causes the lagging of GPU kernels on rank-1.
Although the communication kernel on rank-0 is issued without stalling, it simply waits for the one on rank-1, ultimately causing the overall training speed to decline.
\protect\circlenumber{4}--2 in \autoref{fig:aggregated-metric} shows an example of unnecessary GPU synchronization introduced by the developers from the algorithm teams.
As all ranks wait for the completion of communication kernels, the kernel issue of follow-up kernels is stalled and not overlapped with GPU computation.
When such unnecessary synchronization occurs repeatedly across the model’s forward pass, it ultimately results in a regression of the training speed.

Originally, detecting these anomalies of kernel-issue stall requires investigating the aggregated timeline with much human effort.
Faced with this issue, \sysname{} proposes a new metric, named issue latency distribution, for diagnosing this issue without human intervention.
Kernel-issue latency is defined as the time elapsed between the kernel’s issue timestamp and the start timestamp of its execution on the GPU.
Based on our observation of regression due to kernel-issue stall, the kernel-issue latencies of unhealthy training jobs should be much shorter than those of a healthy training job.

Ahead of deployment, \sysname{} learns healthy kernel issue distributions from historical data for each backend type and cluster scale.
\sysname{} use the maximum Wasserstein distance~\cite{ramdas2017wasserstein} between these healthy distributions as a threshold.
At runtime, Flare compares the collected data against the healthy distribution by computing the Wasserstein distance between them.
A warning is triggered when the distance exceeds the learned threshold.

\paragraph{\protect\circlenumber{5} Void percentage for other un-covered operations.}
While the tracing daemon only instruments the critical operators, inter-step CPU operations and minority GPU kernels both manifest as empty time slots in the visualized timeline, as shown in \autoref{fig:aggregated-metric}.
Consequently, \sysname{} introduces a metric, termed the void percentage, to identify slowdowns caused by these factors.

As for inter-step CPU operations, as depicted by \circlenumber{5}–2 in \autoref{fig:aggregated-metric}, \sysname{} measures the latency between the last kernel preceding the dataloader and the first kernel following the same dataloader. \sysname{} then computes the void percentage for inter-step CPU operations using the following equation:

{
\footnotesize
\begin{equation}
V_{inter} = T_{inter}\ /\ T_{step} 
\end{equation}
}
where $T_{inter}$ represents the latency associated with inter-step CPU operations, and $T_{step}$ denotes the total latency of the training step.

As for minority GPU kernels, as shown by \circlenumber{5}–1 in \autoref{fig:aggregated-metric}, \sysname{} first automatically detects empty slots where GPU kernels are launched but remain un-executed. These empty slots signify that the GPUs are occupied by kernels outside the scope of \sysname{}’s tracing mechanism. \sysname{} subsequently accumulates these slots for each training step and computes the void percentage using the following equation:

{
\footnotesize
\begin{equation}
V_{minority} = T_{minority}\ /\ (T_{step} - T_{inter})
\end{equation}
}
where $T_{minority}$ is the latency of all minority GPU kernels.

When the void percentages ($V_{inter}$ and $V_{minority}$) surpass the predefined thresholds for a specific parallel backend, \sysname{} annotates the training job with potential regressions attributed to inter-step CPU operations or minority GPU kernels. 

\subsubsection{Diagnosing Root Causes for Fail-slows}

While changes in training throughput between steps of the same job indicate fail-slows, \sysname{} cannot directly diagnose the specific factors behind these slowdowns.
Following prior work, \sysname{} investigates the root causes using two micro metrics mentioned above: FLOPS and bandwidth.

By comparing the FLOPS of identical kernels across different ranks, \sysname{} diagnoses GPUs that exhibit poor computational performance, often caused by issues like GPU underclocking.
Machines affected by GPU underclocking are then routed to the operations team for isolation.
The captured communication bandwidth is compared with offline profiled data.
If low-bandwidth communication is detected, \sysname{} conducts a communication test using binary search to pinpoint machines experiencing issues such as network congestion.
These slowdowns are then identified and routed to the operations team for resolution.

\subsubsection{Diagnosing Root Causes for Regressions}
Pinpointing the root causes of regressions in computation and communication kernels is straightforward.
\sysname{} directly forwards the traced data, such as input shape and layout, to the infrastructure team for resolution.
When \sysname{} detects regressions related to kernel-issue stalls, it further narrows down the root causes by analyzing the invocation of relevant Python APIs recorded by the tracing daemon.
Specifically, \sysname{} checks for APIs such as Python GC invoked just before communication kernels with abnormal issue distributions.
The same approach applies to regressions detected by void percentage.
If relevant APIs are found, \sysname{} directly pinpoints the potential root causes and first routes them to the corresponding algorithm teams.
If the algorithm teams cannot resolve them independently, or if no relevant Python APIs are found, \sysname{} forwards the regressions to the infrastructure team for further investigation.
In this case, both algorithm and infrastructure teams validates and resolves regressions with the root causes already narrowed down.



%% file: content/4-implementation.tex


%% file: content/5-evaluation.tex
\section{Evaluation}
\sysname{} consists of $11.4K$ lines of code: 2,197 lines in Python, 7,447 lines in C++, and the rest in supporting code to facilitate its usage.
Beyond the automatic diagnostic workflow described above, \sysname{} also provides rich information to assist manual optimizations, e.g., visualized distributed training timeline.
In this section, we present experiments to demonstrate \sysname{}’s effectiveness from various perspectives.

\subsection{Functionality Comparison}


Since most related works~\cite{jiangMegaScaleScaling,dongEnhancingLargescale,wuGREYHOUNDHunting} are closed-source or only partially open-sourced, we begin with a functionality comparison between \sysname{} and them, followed by in-depth evaluations of \sysname{}.
\autoref{tab:overall-comparison} summarizes the comparison.

Firstly, in terms of tracing mechanisms, only MegaScale and \sysname{} provide full-stack tracing.
Other works, such as C4D, Greyhound, and Holmes, trace only communication or computation kernels.
However, due to different design assumptions, \sysname{} ensures backend extensibility, whereas MegaScale requires intrusive modifications to backend codebases, e.g., patching FSDP before tracing.

Secondly, in diagnostics, only \sysname{} offers automated detection of performance regression anomalies, mainly from algorithm and infrastructure teams, enabled by the proposed micro metrics.
While MegaScale supports full-stack tracing, it only provides distributed visualization for manual regression investigation.
C4D, Greyhound, and Holmes diagnose only errors and fail-slows.
Although \sysname{} also detects these anomalies, it does not enhance their diagnostic mechanisms beyond providing intra-kernel inspection for fast communication hang-error diagnosis.

In a nutshell, \sysname{} is the first diagnostic framework to combine full-stack tracing with backend extensibility and provide automated regression anomaly diagnostics.

\begin{table}[]
\scriptsize
\centering
\setlength{\tabcolsep}{1pt}
\caption{Functionality comparison between \sysname{} and other existing works. ``Comm.'' denotes communication.}
\vspace{-3mm}
\label{tab:overall-comparison}
\begin{tabular}{c|c|c|c|c|c}
\hline
\textbf{Category}                    & \textbf{Features}                                                                                & \textbf{MegaScale}     & \textbf{C4D}           & \textbf{Greyhound}     & \textbf{Flare}                                                          \\ \hline
\multirow{5}{*}{\textbf{\begin{tabular}[c]{@{}c@{}}User\\ experience\end{tabular}}}    & \textbf{Full-stack tracing}                                                                      & \ding{51} & \ding{55} & \ding{55} & \textcolor{LightCayenne}{\ding{51}} \\ \cline{2-6} 
                                     & \textbf{LightCayenneBackend-extensible}                                                                      & \ding{55} & \ding{51} & \ding{51} & \textcolor{LightCayenne}{\ding{51}} \\ \cline{2-6}
                                     & \textbf{Easy-to-play interfaces}                                                                 & \ding{51} & \ding{55} & \ding{55} & \textcolor{LightCayenne}{\ding{51}} \\ \cline{2-6} 
                                     & \textbf{\begin{tabular}[c]{@{}c@{}}Automated diagnostics\\ with aggregated metrics\end{tabular}} & \ding{55} & \ding{55} & \ding{55} & \textcolor{LightCayenne}{\ding{51}} \\ \cline{2-6} 
                                     & Distributed visualization                                                                        & \ding{51} & \ding{55} & \ding{55} & \textcolor{LightCayenne}{\ding{51}} \\ \hline
\multirow{2}{*}{\textbf{Hang error}} & Non-comm. hang                                                                                   & \ding{51} & \ding{51} & \ding{55} & \textcolor{LightCayenne}{\ding{51}} \\ \cline{2-6} 
                                     & Comm. hang                                                                                       & $\geq 30min$           & $\geq 30min$           & \ding{55} & \textcolor{LightCayenne}{$\leq 5min$} \\ \hline
\multirow{5}{*}{\textbf{Slowdown}}   & Critical kernels                                                                                 & \ding{51} & \ding{55} & \ding{51} & \textcolor{LightCayenne}{\ding{51}} \\ \cline{2-6} 
                                     & \begin{tabular}[c]{@{}c@{}}Overlapping of \\ Comp. and Comm.\end{tabular}                        & \ding{51} & \ding{55} & \ding{55} & \textcolor{LightCayenne}{\ding{51}} \\ \cline{2-6} 
                                     & Comm. kernels                                                                                    & \ding{51} & \ding{51} & \ding{51} & \textcolor{LightCayenne}{\ding{51}} \\ \cline{2-6} 
                                     & \textbf{Kernel-issue stall}                                                                      & Only GC                & \ding{55} & \ding{55} & \textcolor{LightCayenne}{\ding{51}} \\ \cline{2-6} 
                                     & \textbf{Less critical operations}                                                               & \ding{55} & \ding{55} & \ding{55} & \textcolor{LightCayenne}{\ding{51}} \\ \hline
\end{tabular}
\vspace{-3mm}
\end{table}

\subsection{Runtime Overhead}
\label{sec:eval:overhead}
\begin{figure}
    \centering
    \includegraphics[width=.9\linewidth]{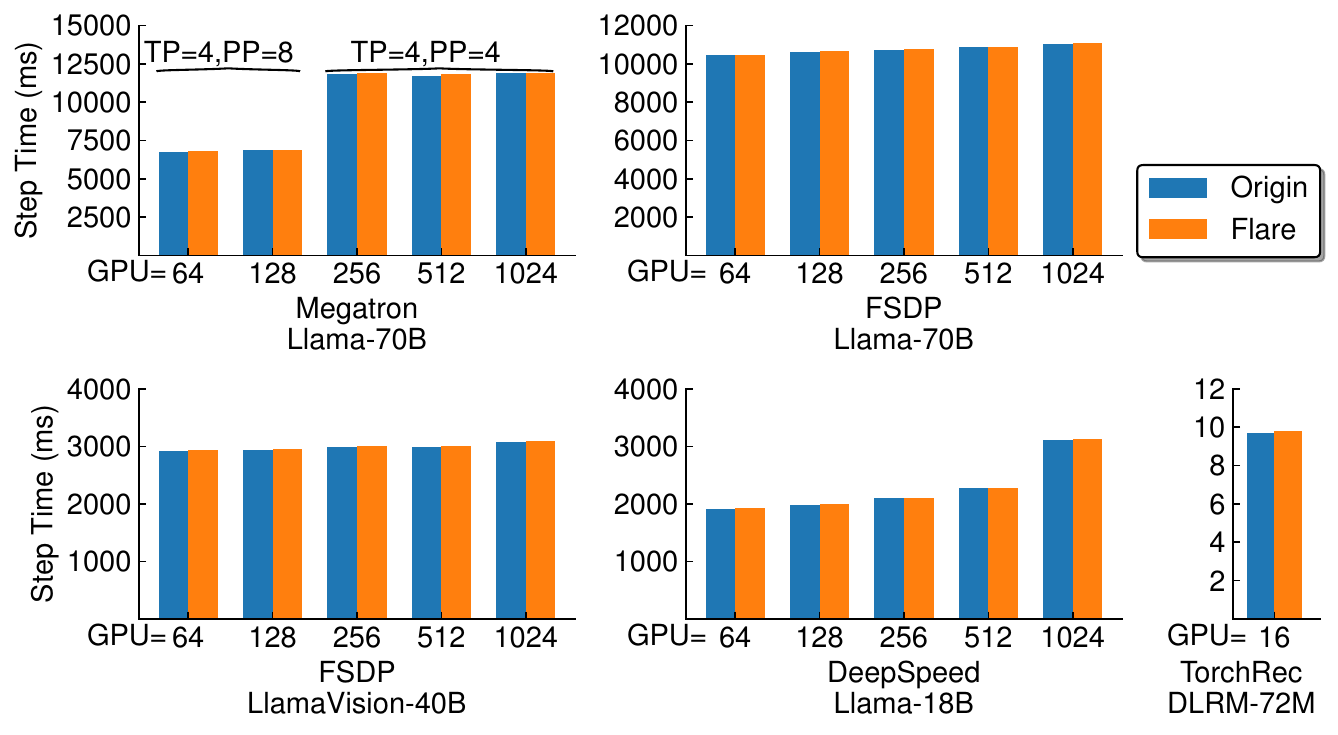}
    \vspace{-3mm}
    \caption{Runtime overhead in terms of latency with various models, backends, and number of GPUs.}
    \label{fig:eval:latency-overhead}
    \vspace{-5mm}
\end{figure}
We evaluate \sysname{} on four parallel backends: Megatron~\cite{shoeybiMegatronLMTraining}, FSDP~\cite{zhaoPyTorchFSDP}, DeepSpeed~\cite{rajbhandariZeROMemory}, and TorchRec~\cite{ivchenkoTorchRecPyTorch}. Among these, Megatron, FSDP, and DeepSpeed are widely used for LLM training, while TorchRec is employed for training large recommendation models within Ant Group. Four models are benchmarked, spanning language, vision, and recommendation tasks: two large language models (Llama 18B and 70B), one large vision model (Llama Vision 40B), and one recommendation model (DLRM 72M).

The latency overhead experiment is conducted on 1,024 H800 GPUs deployed across 128 servers with RoCE connectivity. 
We directly compare \sysname{} with the original execution, as it serves as the lower bound.
As shown, \sysname{} incurs a latency overhead of $0.43\%$ for three LLM training backends and $1.02\%$ for TorchRec.
Across other GPUs like A100, \sysname{} shows consistent results.

We also compared the runtime overhead of \sysname{} with existing works~\cite{jiangMegaScaleScaling,wuGREYHOUNDHunting}.
Compared with MegaScale, \sysname{} incurs similar runtime overhead, as both selectively trace key code segments.
The main difference is that \sysname{} avoids intruding into backend codebases, achieving better backend extensibility. Originally, Greyhound only traces start timestamps of communication kernels.
We extend Greyhound’s tracing mechanism to support full-stack tracing.
In this case, Greyhound incurs unacceptable overhead, reaching $35\%$ latency with Llama-8B trained on just 8 GPUs.
This is because Greyhound's tracing mechanism is tailored for fail-slows and does not cover regressions.


The memory overhead experiment is conducted on two setups, which are 16 A100 GPUs on 2 nodes and 1536 H800 GPUs on 192 nodes.
The testbed slightly differs from that used in the latency overhead evaluation, as we conducted the experiments on available idle GPUs due to heavy load on our training cluster.
We compare \sysname{} with \texttt{Torch Full}, \texttt{Torch w/o Stack}, \texttt{Torch w/o Layout\&Stack}, and \sysname{}. \texttt{Torch w/o Stack} refers to using the PyTorch builtin profiler with stack tracing disabled, while \texttt{Torch w/o Layout\&Stack} further disables matrix layout tracing.

\begin{figure}
    \centering
    \includegraphics[width=0.76\linewidth]{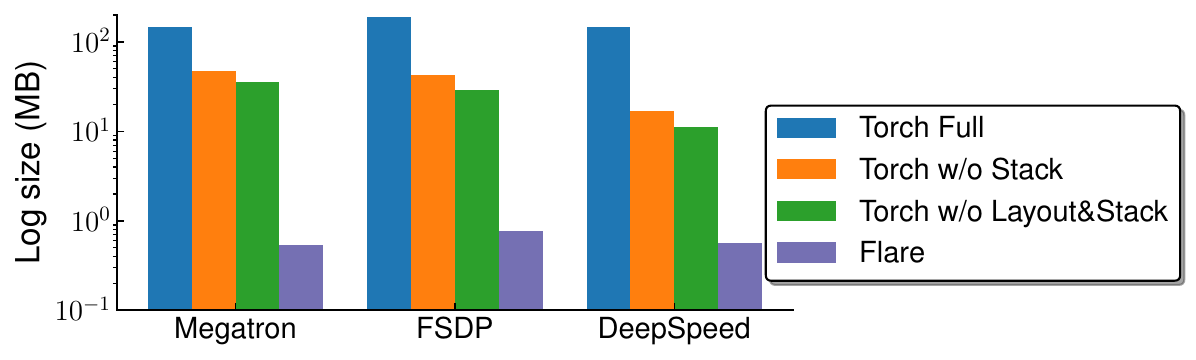}
    \vspace{-2mm}
    \caption{Memory consumption of dumped logs per GPU per step using the PyTorch profiler and \sysname{} while training a Llama-70B model on 16 A100 GPUs.}
    \label{fig:log-overhead}
    \vspace{-4mm}
\end{figure}

\autoref{fig:log-overhead} shows the memory overhead results on 16 A100 GPUs. \sysname{} consumes only $0.39\%$, $1.76\%$, and $2.48\%$ of memory overhead for the respective configurations of PyTorch profiler. \sysname{} generates a maximum of $0.78\text{MB}$ of tracing logs per GPU. Besides, in a real-world Llama-20B training job on 1536 H800 GPUs, \sysname{} generated only a $1.5\text{MB}$ tracing log per GPU.
TorchRec is omitted from this experiment, as monitoring a recommendation model generates minimal logs.
From the above results, \sysname{} consistently maintains an extremely low runtime overhead in terms of both latency and memory.
The lightweight selective tracing facilitates \sysname{}'s deployment within our training cluster, serving as a diagnostic framework for diverse training jobs.

\subsection{Effectiveness of Intra-kernel Inspecting}

We evaluate the intra-kernel inspecting mechanism on 16 A100 GPUs across two servers with RoCE connectivity. Given that most communication kernels are ring-based, this experiment focuses on evaluating ring-allreduce.
We customize the training script composed solely of communication kernels, with one GPU intentionally suspended to simulate a hang error caused by communication issues. 

\begin{figure}
    \centering
    \includegraphics[width=0.58\linewidth]{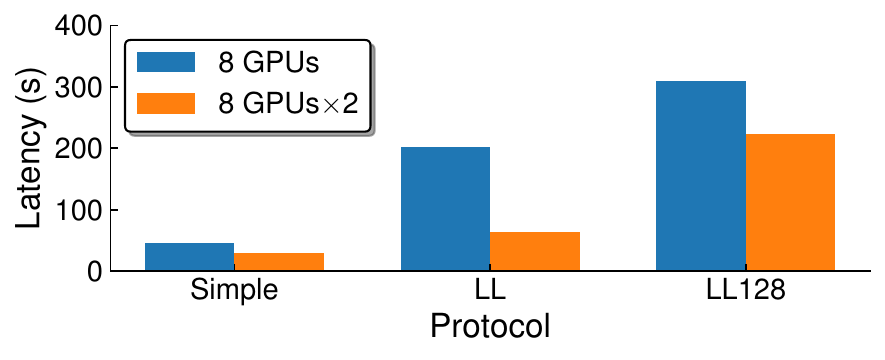}
    \vspace{-3mm}
    \caption{Latency for pinpointing the erroneous GPUs causing a hang error in ring-allreduce with different protocols.}
    \label{fig:eval:intra-kernel}
    \vspace{-5mm}
\end{figure}

\autoref{fig:eval:intra-kernel} illustrates the pinpointing latencies for intra- and inter-server communication.
The figure presents the latency results for three communication protocols~\cite{NVIDIA_NCCL_User_Guide} and cross-node configurations.
As shown, \sysname{} requires $29.4\text{\textasciitilde}309.2s$ to detect erroneous GPUs across different scenarios. Among the protocols, \sysname{} performs best when the SIMPLE protocol is used for communication. This is because, with the SIMPLE protocol, \sysname{} only needs to scan the first thread of each thread block to check the steps, whereas the other two protocols require scanning the entire thread block.

When comparing intra-server and inter-server results, \sysname{} performs better when the ring-allreduce operation spans multiple servers. This is because intra-server GPUs are connected via NVLink, whereas inter-server GPUs communicate through NICs. Communication kernels launched over NICs involve fewer thread blocks, as NICs have fewer internal links compared to NVLink. As a result, \sysname{} scans fewer thread blocks for error diagnosis in inter-server scenarios.

In summary, the intra-kernel inspecting mechanism can detect erroneous GPUs in a maximum of $309.2s$.
Notably, as the complexity of intra-kernel inspecting is $O(1)$, these results remain constant.

\begin{figure}
    \centering
    \includegraphics[width=.76\linewidth]{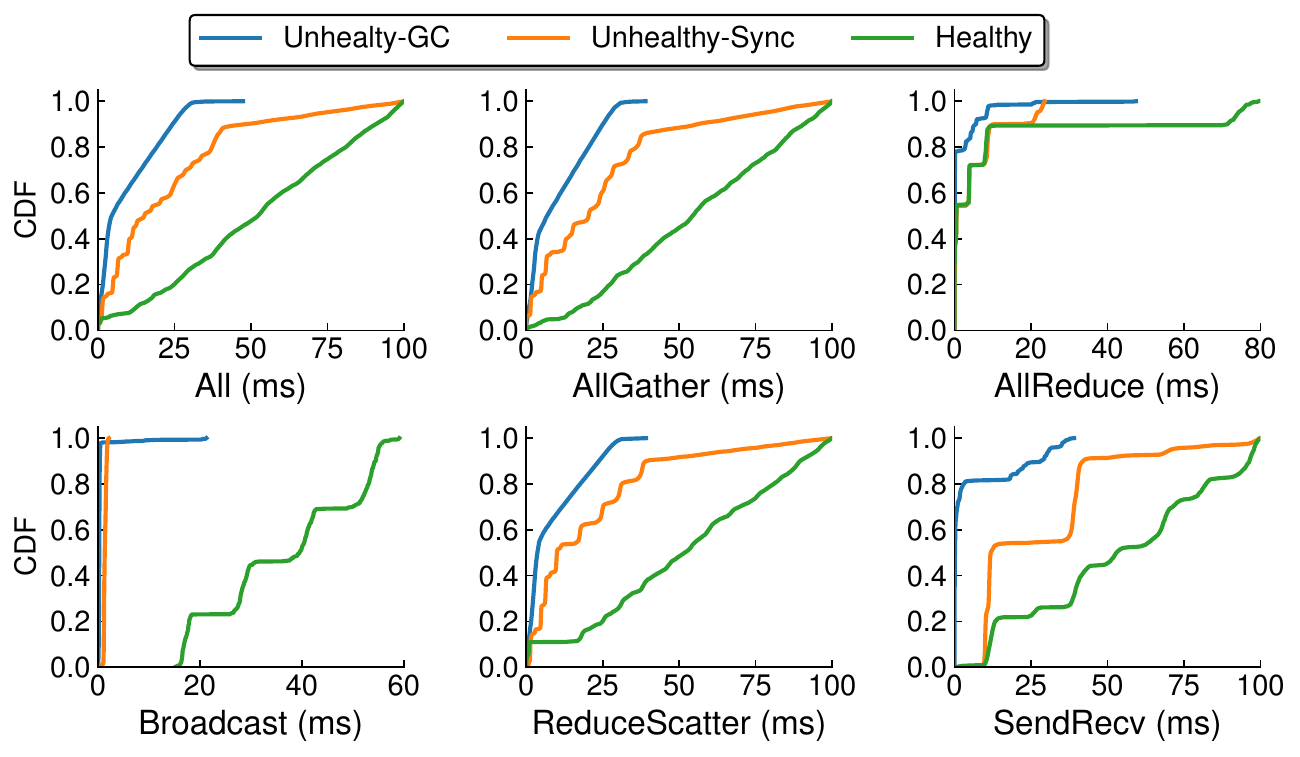}
    \vspace{-2mm}
    \caption{Issue distribution for a Llama-20B model trained with Megatron and 256 GPUs, including the overall CDF and the CDFs for each type of communication kernel, respectively.}
    \label{fig:eval:issue-distribution}
    \vspace{-4mm}
\end{figure}

\subsection{Effectiveness of Issue Latency Distribution}
\label{sec:eval:issue}

In this experiment, we evaluate the issue latency distribution using Llama-20B running on 256 H800 GPUs across 32 servers connected via RoCE. \autoref{fig:eval:issue-distribution} illustrates the issue latency distribution for all communication kernels in the \texttt{Unhealthy-GC}, \texttt{Unhealthy-Sync}, and \texttt{Healthy} scenarios. In the \texttt{Unhealthy-GC} scenario, GC is implicitly triggered by the Python runtime. In the \texttt{Unhealthy-Sync} scenario, an unintended GPU synchronization call is added within the transformer block, leading to repetitive GPU synchronizations during the model’s forward pass. In the \texttt{Healthy} scenario, GC is efficiently managed by the parallel backend, and no unnecessary synchronizations are introduced.


As shown in the figure, the issue latency distribution patterns align with our claim in \S\ref{sec:diagnose:issue}. The issue latency CDF of a healthy LLM training job increases linearly, whereas the issue latency CDFs for \texttt{Unhealthy-GC} and \texttt{Unhealthy-Sync} exhibit a much steeper rise. This is because the issue latencies of different ranks in the healthy scenario are solely influenced by the collective communication operator, resulting in a uniform distribution. In contrast, in the cases of \texttt{Unhealthy-GC} and \texttt{Unhealthy-Sync}, while some ranks are affected, their latencies become very short due to the delayed start of the issue time.
Since both GC and GPU synchronizations span the entire model forward pass, all communication kernels are affected, as illustrated in \autoref{fig:eval:issue-distribution}. Furthermore, each training process triggers GC independently, and the GC operation for a single process is more time-consuming than GPU synchronization. Consequently, the issue latency distribution for \texttt{Unhealthy-GC} is worse than that of \texttt{Unhealthy-Sync}.


\section{Deployment \& Case Studies}
\label{sec:case_study}

\subsection{Cluster-wide Deployment}

\sysname{} has been deployed and running continuously in a training cluster with 6,000 GPUs for over eight months.
During this period, it is responsible for monitoring, detecting, and diagnosing training jobs for various deep learning models, especially large-scale distributed LLM training.


\begin{table}
\centering
\scriptsize
\setlength{\tabcolsep}{4pt}
\caption{Typical errors detected by \sysname{}.}
    \vspace{-2mm}
\label{tb:case:errors}
\begin{tabular}{c|c|c|c}
\hline
\textbf{Taxonomy} & \textbf{Details} & \textbf{Numbers} & \textbf{Mechanism} \\ \hline
\multirow{2}{*}{\begin{tabular}[c]{@{}c@{}}OS\\ errors\end{tabular}} & Checkpoint storage & 10 & \multirow{4}{*}{\begin{tabular}[c]{@{}c@{}}Stack\\ analysis\end{tabular}} \\ \cline{2-3}
 & OS crash & 1 &  \\ \cline{1-3}
\multirow{3}{*}{\begin{tabular}[c]{@{}c@{}}GPU\\ errors\end{tabular}} & GPU Driver & 26 &  \\ \cline{2-3}
 & Faulty GPU (Unknown) & 37 &  \\ \cline{2-4} 
 & \multirow{2}{*}{NCCL hang} & \multirow{2}{*}{36} & \multirow{3}{*}{\begin{tabular}[c]{@{}c@{}}Intra-kernel\\ tracing\end{tabular}} \\ \cline{1-1}
\multirow{2}{*}{\begin{tabular}[c]{@{}c@{}}Network\\ errors\end{tabular}} &  &  &  \\ \cline{2-3}
 & RoCE issue & 17 &  \\ \hline
\end{tabular}
\vspace{-5mm}
\end{table}

\subsection{Errors}
\autoref{tb:case:errors} presents a subset of error and fail-slow anomalies detected by \sysname{}.
As shown in the table, \sysname{} effectively identifies OS- and hardware-related anomalies, including crashes, and hangs.
While these issues are typically conspicuous and could be detected by existing methods based on noticeable training interruptions~\cite{shoeybiMegatronLMTraining,wuGREYHOUNDHunting}, \sysname{}’s novelty lies in providing richer error information through techniques like intra-kernel inspecting.
Such runtime information helps ease and accelerate the attribution process of low-level issues for operations teams.

\subsection{Fail-slows and Performance Regressions}

\autoref{tb:case:slowdowns} summarizes fail-slow and regression slowdowns diagnosed by \sysname{} using its aggregated metrics.
We omit evaluation of \sysname{} in fail-slow diagnostics, as it employs similar mechanisms to prior works~\cite{dongEnhancingLargescale,wuGREYHOUNDHunting}.

To evaluate regression diagnostics, we collect 113 real-world training jobs submitted within a week in a 6,000-GPU cluster.
During this period, the cluster maintained an average GPU usage of 80\%.
The jobs included popular LLMs, multi-modal LLMs, and recommendation models.
We first diagnose them with \sysname{}, then validate results against human-labeled ground truth.
It successfully diagnosed 9 true regressions based on issue latency distribution and void percentage, with only 2 false ones.
This corresponds to a false positive rate of $1.9\%$ and a true positive diagnostic accuracy of $81.8\%$.

We further investigated the 2 false positives.
The first was a multi-modal LLM trained with FSDP, where input data contained images of varying resolutions.
This caused imbalanced computation across ranks, leading to abnormal issue latency distribution.
The second was a recommendation model using CPU-based embeddings.
Its void percentage was higher than that of GPU-based ones, causing misclassification.

Although these 2 cases were falsely diagnosed as regressions, they highlight opportunities to refine \sysname{}’s micro-metrics with historical data.
For example, relaxing latency distribution thresholds for imbalanced multi-modal inputs and adjusting void percentage thresholds for CPU-based embedding models could reduce such false positives.
After these refinements, \sysname{} no longer misclassifies these job types in real-world cluster deployments.

Now, we present several typical diagnosed regressions.


\begin{table}
\centering
\scriptsize
\setlength{\tabcolsep}{4pt}
\caption{Fail-slows and regressions diagnosed by \sysname{}, with ``\textbf{Details}'' showing training job specifics and associated MFU decline.
We show the diagnosed regression anomalies by \sysname{} in bold.
}
\vspace{-1mm}
\label{tb:case:slowdowns}
\begin{tabular}{c|c|c}
\hline
\textbf{Metric} & \textbf{Attribution} & \textbf{Details} \\ \hline
\multirow{2}{*}{FLOPS} & GPU underclocking & 480 GPUs, Llama-65B, $14\%\downarrow$ \\ \cline{2-3} 
 & \textbf{Backend migration} & \textbf{1856 GPUs, Llama-80B, $33.3\%\downarrow$} \\ \hline
\multirow{3}{*}{Bandwidth} & \begin{tabular}[c]{@{}c@{}}Network jitter with\\ increased CRC\end{tabular} & 928GPUs, Llama-65B, $10\text{\textasciitilde}20\%\downarrow$ \\ \cline{2-3} 
 & \begin{tabular}[c]{@{}c@{}}Down of\\ GDR module\end{tabular} & \begin{tabular}[c]{@{}c@{}}32GPUs, Llama-10B, $80\%\downarrow$\\ 128GPUs, Llama-10B, $62.5\%\downarrow$\\ ...\end{tabular} \\ \cline{2-3} 
 & \begin{tabular}[c]{@{}c@{}}Host-side hugepage\\ caused high sysload\end{tabular} & 128GPUs, LlamaVision-11B, $20\%\downarrow$ \\ \hline
\multirow{4}{*}{\begin{tabular}[c]{@{}c@{}}Issue latency\\ distribution\end{tabular}} & \textbf{Python GC} & \begin{tabular}[c]{@{}c@{}}2048GPUs, Llama-80B, $10\%\downarrow$\\ 280GPUs, LlamaVision-11B, $60\%\downarrow$\\ ...\end{tabular} \\ \cline{2-3} 
 & \textbf{\begin{tabular}[c]{@{}c@{}}Unnecessary\\ GPU Sync\end{tabular}} & \textbf{\begin{tabular}[c]{@{}c@{}}256GPUs, Llama-20B, $2.66\%\downarrow$\\ ...\end{tabular}} \\ \cline{2-3} 
 & \textbf{Package chcecking} & \textbf{280GPUs, LlamaVision-20B, $30\%\downarrow$} \\ \cline{2-3} 
 & \multirow{2}{*}{\textbf{\begin{tabular}[c]{@{}c@{}}Frequent GPU\\ mem. management\end{tabular}}} & \multirow{2}{*}{\textbf{1344GPUs, Llama-176B, $19\%\downarrow$}} \\ \cline{1-1}
\multirow{2}{*}{\begin{tabular}[c]{@{}c@{}}Void\\ percentage\end{tabular}} &  &  \\ \cline{2-3} 
 & \textbf{Dataloader} & \textbf{512GPUs, Llama-80B, $41\%\downarrow$} \\ \hline
\end{tabular}
\vspace{-5mm}
\end{table}

\subsubsection{Case-1: Towards Stall-free Kernel Issuing}
\label{sec:case:stall-free}


Kernel-issue stalls are among the most frequent causes of slowdowns encountered in a training cluster not dedicated exclusively to a single pre-training task.
While Python runtime GC is well-known and now carefully managed by the parallel backend in most cases, most encountered kernel-issue stalls arise from code introduced by algorithm teams to enhance LLM performance in downstream tasks.

A typical case encountered by \sysname{} is a training job of Llama-20B running on 256 H800 GPUs. The developer from the algorithm team mistakenly enables the timer provided by Megatron for performance profiling of several key code segments. This profiling incurs kernel-issue stalls because it requires GPU synchronizations to obtain accurate timestamps.

Although no significant regression was observed in training throughput, \sysname{} successfully detects the abnormal issue latency distribution.
After removing these unnecessary synchronizations by disabling the timer, the MFU of the training job persistently improves from $41.4\%$ to $42.5\%$, representing a $2.66\%$ increase.
Such regression is minor, which could not be directly observed through macro metrics like training throughput in a large training cluster with a lot of noise.
However, there is an obvious drift in micro metrics like issue latency distribution, enabling \sysname{} to detect such obsecured regression anomalies.

In addition to the above two cases, \sysname{} also detects other kernel-issue stalls, such as unnecessary package version checking, frequent CUDA memory management within the PyTorch runtime, and others.

\subsubsection{Case-2: Migration between Backends}

Different parallel backends are suited to varying hardware conditions.
In this context, algorithm teams may migrate an LLM between backends to meet their specific demands. 
However, this migration process can potentially introduce regressions.
A typical scenario encountered by \sysname{} is an anomalous MFU decline when migrating a Llama-like 80B model from FSDP (1888 H800 GPUs) to Megatron (1586 H800 GPUs with a data-parallel degree of 58, pipeline-parallel degree of 8, and tensor-parallel degree of 4). This regression specifically stems from a matrix layout change.
The weight dimension of the LLM’s FFN (feed forward network) layer, initially configured as $[8192 \times 33936]$ during training on FSDP, changes to $[8192 \times 8484]$ after migration to Megatron with a tensor parallelism degree of 4.

After migration to Megatron, this operator exhibits significantly lower FLOPS due to smaller batch size and the unfavorable 8484 layout for Tensor Cores, which require alignment to 128 bytes.
In contrast, the dimension 33936 and larger batch size on FSDP meet this alignment requirement.

\begin{figure}
    \centering
    \includegraphics[width=0.64\linewidth]{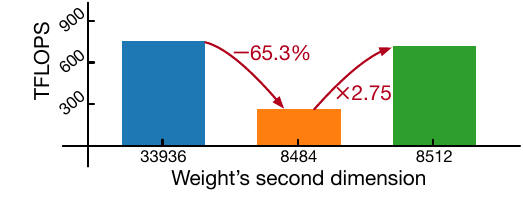}
    \vspace{-3mm}
    \caption{The change in computation TFLOPS when migrating Llama-80B from FSDP to Megatron.}
    \label{fig:layout}
    \vspace{-2mm}
\end{figure}

While the algorithm team does not report this regression, \sysname{} detected and routed it to our infrastructure team.
Following \sysname{}’s diagnosis, our infrastructure team customizes a kernel that pads 8484 to 8512. \autoref{fig:layout} illustrates the FLOPS of the same operator before migration, after migration, and post-optimization guided by \sysname{}.
As shown, the operator experiences a $65.3\%$ decline in FLOPS after migration, and \sysname{} successfully facilitates the performance diagnosis.
From the perspective of the training job, the overall MFU increases from $27\%$ to $36\%$, reflecting a $33.3\%$ improvement.


\subsubsection{Case-3: New Algorithms and Data}
Algorithm teams continually strive to enhance model performance by modifying the LLM architecture and incorporating new training data. However, this often introduces regressions.
Firstly, algorithm teams generally modify position embeddings (PE), activation functions (ACT), and normalization operators (NORM), while preserving the core structure of the transformer. \autoref{tb:void-minority} illustrates the detected changes in $V_{minority}$ caused by these operator modifications during daily deployments. In this table, the parallel backend is Megatron. The \texttt{Healthy} column represents a fully optimized training job, whereas the \texttt{-PE} column reflects changes in modifying the position embeddings. Similarly, the other columns correspond to modifications of the respective operators.

Since these modified operators are less critical and not instrumented by \sysname{}, $V_{minority}$ increases proportionally with their computational complexity.
Our infrastructure team leverages \sysname{}’s detection of high $V_{minority}$ to develop targeted kernel implementations.
Once optimized through techniques like kernel fusion, the job’s $V_{minority}$ returns to a normal level. In this process, \sysname{} eliminates the need for manual identification, thereby accelerating anomaly diagnosis.


Secondly, newly filtered data is continuously incorporated into the LLM for training. \sysname{} has successfully diagnosed anomalies arising from variance in the training data.
In one specific case, the algorithm team attempts to train Llama-80B with data containing a sequence length of 64k, while the original training script is for sequence lengths of 4k. \sysname{} identifies a significant anomalous decline in MFU ($41\%$) on 512 H800 GPUs, accompanied by an increase in $V_{inter}$.

After routing this anomaly with the root cause identified in the dataloader to the algorithm team, they validated that the regression was due to the attention mask generation process within the dataloader.
When the sequence length is short, the latency incurred by mask generation is minimal. However, the complexity of mask generation scales as $O(L^2)$, where $L$ represents the sequence length. As a result, the dataloader experiences extremely poor performance when the sequence length increases to 64k.


\begin{table}
\scriptsize
\centering
\caption{Changes in detected $V_{minority}$ and normalized TFLOPS when different minority kernels are not optimized.}
\vspace{-2mm}
\label{tb:void-minority}
\begin{tabular}{c|c|c|c|c}
\hline
                        & \textbf{Healthy} & \textbf{-PE} & \textbf{-PE-ACT} & \textbf{-PE-ACT-NORM} \\ \hline
$\mathbf{V_{minority}}$ & $9\%$           & $14\%$       & $15\%$           & $28\%$                \\ \hline
\textbf{N. TFLOPS}      & 1               & 0.95         & 0.93             & 0.83                  \\ \hline
\end{tabular}
\vspace{-5mm}
\end{table}



\section{Experience}
\subsection{Practical Usages}
\label{sec:practical}
By leveraging \sysname{}, algorithm teams can independently identify and resolve regression anomalies caused by inefficient code without requiring intervention from the infrastructure team. With \sysname{}’s lightweight logging, the infrastructure team also collects sufficient runtime data to analyze submitted jobs and independently discover new optimization opportunities. Only regression anomalies that cannot be resolved by algorithm teams are routed to the infrastructure team. Overall, the frequency of collaboration on recurrent regressions, such as unnecessary synchronizations, decreased by 63.5\% within a one-week deployment, accelerating model evolution for algorithm teams and allowing the infrastructure team to focus more on system optimization.

\subsection{Using Historical Data}
Historical traces are essential for enhancing the effectiveness of regression detection.
While runtime data is crucial, it is insufficient on its own for accurate regression diagnosis.
\sysname{} detects issues by comparing real-time data against historical data. E.g., when using issue latency distributions to detect kernel-issue stalls, \sysname{} relies on historical data from specific backends operating on specific hardware in \autoref{fig:eval:issue-distribution}. 
However, regression detection is inherently difficult to ascertain, even with historical data.
Therefore, in practice, we adopt a conservative policy: we report potential regressions and their root causes without directly terminating training jobs.

\subsection{Hardware Extending}
Currently, \sysname{} has supported the extensibility of additional hardware, particularly NPUs dedicated to DL training.
Since \sysname{} directly instruments key code segments at the Python and C++ runtime levels, extending it is straightforward.
We already support internal CUDA-native NPUs, incurring less than $0.5\%$ runtime overhead when running on 450 NPUs.
We also investigate the extensibility of \sysname{}’s intra-kernel inspecting on NPUs.
Our study shows that the methodology is largely extensible, as NPUs also use dedicated hardware cores for cross-device communication.
Inspecting their register status provides additional information for the operations team to address hang-error anomalies, but this requires additional engineering effort.

\subsection{Scopes \& Limitations}

\sysname{} operates within certain scopes and has limitations.

\textit{Firstly,
\sysname{} targets large-scale private GPU clusters for LLM training and relies on historical data to calibrate its diagnostic metrics.}
This assumption is reasonable.
Accessing historical data is practical in a private GPU clusters.
In our internal cluster, we also profiled typical LLMs and parallel backends at different scales to obtain ground-truth data for building the diagnostic mechanism.

\textit{Secondly,
\sysname{} cannot detect regressions in training jobs with major architecture updates, e.g., switching from a dense LLM to a sparse MoE LLM.}
In such cases, \sysname{} must collect new historical data to refine its diagnostic engine.
This requirement is usually reasonable, as \sysname{} is designed to relieve the infrastructure team from repetitive efforts on recurrent regression anomalies, allowing them to focus instead on optimizing LLMs with new architectures.

%% file: content/6-related_work.tex
\section{Related Work}
\paragraph{Anomaly diagnosis in distributed training. }
Anomaly diagnosis in large-scale distributed deep learning training tasks has consistently been a hot research focus\cite{jiangMegaScaleScaling, wuGREYHOUNDHunting, dongEnhancingLargescale, peng2018optimus, haider2011fault, chen2020elastic}. Megascale~\cite{jiangMegaScaleScaling} only focuses on LLM training tasks based on Megatron-LM. It identifies network-related hardware and software issues in the training process by conducting intra-host network tests and NCCL tests. Greyhound~\cite{wuGREYHOUNDHunting} detects prolonged iterations using the Bayesian Online Change-Point Detection algorithm.
C4D~\cite{dongEnhancingLargescale} modifies the Collective Communication Library to collect message statistics, such as sizes and durations of transfers, to identify the performance bottlenecks. However, these approaches primarily address communication-related performance issues and fail to encompass the anomalies across the LLM training stack.
Additionally, they depend heavily on a single parallel backend, limiting their generality.

\paragraph{Anomaly diagnosis in large-scale datacenter. }

Many research efforts\cite{bergerTriangulatingPython, wangDiagnosingApplicationnetwork,yangAAsclepiusMonitoring , han2021depth, wangZeroOverhead} focus on anomaly diagnosis at different levels of the datacenter, including runtime, network, and storage.
AND\cite{wangDiagnosingApplicationnetwork} is a unified application-network diagnosing system that leverages a single metric, TCP retransmissions (TCP retx), to identify network anomalies in cloud-native scenarios.
AAsclepius\cite{yangAAsclepiusMonitoring} proposes a PathDebugging technique to trace fault linkages between the middle network and autonomous systems.
Researchers~\cite{han2021depth} from Alibaba analyze four key factors that impact SSD failure correlations: drive models, lithography, age, and capacity.
As these works address anomaly problems in specific scenarios, they are unable to resolve the challenges targeted by \sysname{} in large-scale distributed LLM training.

%% file: content/7-conclusion.tex
\section{Conclusion}

In this paper, we introduce \sysname{}, a real-time diagnostic framework for LLM training. By addressing challenges such as lightweight long-term monitoring, root cause detection, and backend extensibility, \sysname{} provides a comprehensive solution for anomaly diagnostics across large-scale GPU clusters. With its novel intra-kernel inspecting mechanism and holistic aggregated metrics, \sysname{} not only reduces diagnostic complexity but also improves the detection and resolution of non-obvious slowdowns and errors. Its deployment in real-world training clusters, spanning over 6,000 GPUs, demonstrates its efficacy in diagnosing distributed LLM training.

%% file: references/others.bib
@inproceedings{chen2020elastic,
  title={Elastic parameter server load distribution in deep learning clusters},
  author={Chen, Yangrui and Peng, Yanghua and Bao, Yixin and Wu, Chuan and Zhu, Yibo and Guo, Chuanxiong},
  booktitle={Proceedings of the 11th ACM Symposium on Cloud Computing},
  pages={507--521},
  year={2020}
}

@inproceedings{haider2011fault,
  title={Fault tolerance in distributed paradigms},
  author={Haider, Sajjad and Ansari, Naveed Riaz and Akbar, Muhammad and Perwez, Mohammad Raza and Ghori, KM},
  booktitle={In2011 International Conference on Computer Communication and Management, Proc. of CSIT},
  volume={5},
  year={2011}
}

@inproceedings{peng2018optimus,
  title={Optimus: an efficient dynamic resource scheduler for deep learning clusters},
  author={Peng, Yanghua and Bao, Yixin and Chen, Yangrui and Wu, Chuan and Guo, Chuanxiong},
  booktitle={Proceedings of the Thirteenth EuroSys Conference},
  pages={1--14},
  year={2018}
}

@inproceedings{han2021depth,
  title={An In-Depth Study of Correlated Failures in Production SSD-Based Data Centers},
  author={Han, Shujie and Lee, Patrick PC and Xu, Fan and Liu, Yi and He, Cheng and Liu, Jiongzhou},
  booktitle={19th USENIX Conference on File and Storage Technologies (FAST 21)},
  pages={417--429},
  year={2021}
}

@article{deepseekmoe,
  title={Deepseekmoe: Towards ultimate expert specialization in mixture-of-experts language models},
  author={Dai, Damai and Deng, Chengqi and Zhao, Chenggang and Xu, RX and Gao, Huazuo and Chen, Deli and Li, Jiashi and Zeng, Wangding and Yu, Xingkai and Wu, Y and others},
  journal={arXiv preprint arXiv:2401.06066},
  year={2024}
}

@misc{openaio1,
  title = {Learning to Reason with LLMs},
  author = {OpenAI},
  year = {2024},
  howpublished = {\url{https://openai.com/index/learning-to-reason-with-llms/}},
  note = {Accessed: 2024-10-20}
}

@misc{sora,
  title = {Creating video from text},
  author = {OpenAI},
  year = {2024},
  howpublished = {\url{https://openai.com/index/sora/}},
  note = {Accessed: 2024-10-20}
}

@article{rstar,
  title={Mutual reasoning makes smaller llms stronger problem-solvers},
  author={Qi, Zhenting and Ma, Mingyuan and Xu, Jiahang and Zhang, Li Lyna and Yang, Fan and Yang, Mao},
  journal={arXiv preprint arXiv:2408.06195},
  year={2024}
}

@misc{llama3-1,
  title = {Llama 3.1: A New Milestone for Open Large Language Models},
  author = {Hugging Face},
  year = {2024},
  howpublished = {\url{https://huggingface.co/blog/llama31}},
  note = {Accessed: 2024-10-20}
}

@article{vit,
  title={An image is worth 16x16 words: Transformers for image recognition at scale},
  author={Dosovitskiy, Alexey},
  journal={arXiv preprint arXiv:2010.11929},
  year={2020}
}

@misc{xai_colossus_training_cluster,
  title        = {Colossus Training Cluster},
  author       = {xAI},
  year         = {2024},
  howpublished = {\url{https://www.techradar.com/pro/xai-cluster-is-now-the-most-powerful-ai-\\training-system-in-the-world-but-questions\\-remain-over-storage-capacity-power-usage\\-and-why-it-s-actually-called-colossus}}
}

@misc{meta_ai_training_infrastructure,
  title        = {Maintaining large-scale AI capacity at Meta},
  author       = {Meta Engineering Team},
  year         = {2024},
  howpublished = {\url{https://engineering.fb.com/2024/06/12/production-engineering/maintaining-large-scale-ai-capacity-meta}}
}

@misc{nvidia_ampere_architecture,
  title        = {NVIDIA Ampere Architecture},
  author       = {NVIDIA Corporation},
  year         = {2020},
  howpublished = {\url{https://www.nvidia.com/en-us/data-center/ampere-architecture/}}
}

@misc{nvidia_hopper_architecture,
  title        = {NVIDIA Hopper Architecture},
  author       = {NVIDIA Corporation},
  year         = {2022},
  howpublished = {\url{https://www.nvidia.com/en-us/data-center/technologies/hopper-architecture/}}
}

@misc{nvidia_nvswitch,
  title        = {NVIDIA NVSwitch: The World's Highest-Bandwidth On-Node Switch},
  author       = {NVIDIA Corporation},
  year         = {2018},
  howpublished = {\url{https://images.nvidia.com/content/pdf/nvswitch-technical-overview.pdf}}
}

@misc{nvidia_nvl72,
  title        = {NVIDIA GB200 NVL72: HPC \& AI GPU for Data Centers},
  author       = {NVIDIA Corporation},
  year         = {2024},
  howpublished = {\url{https://www.nvidia.com/en-us/data-center/gb200-nvl72/}}
}

@misc{cublas2024,
  title        = {cuBLAS Library User Guide},
  author       = {{NVIDIA Corporation}},
  year         = {2024},
  howpublished = {\url{https://docs.nvidia.com/cuda/archive/12.6.2/cublas/}}
}

@misc{cutlass2024,
  title        = {CUTLASS: CUDA Templates for Linear Algebra Subroutines},
  author       = {{NVIDIA Corporation}},
  year         = {2024},
  howpublished = {\url{https://github.com/NVIDIA/cutlass}}
}

@misc{nccl2024,
  title        = {NCCL: NVIDIA Collective Communication Library},
  author       = {{NVIDIA Corporation}},
  year         = {2024},
  howpublished = {\url{https://developer.nvidia.com/nccl}}
}

@misc{cupti2024,
  title        = {CUDA Profiling Tools Interface (CUPTI) User Guide},
  author       = {{NVIDIA Corporation}},
  year         = {2024},
  howpublished = {\url{https://docs.nvidia.com/cupti/index.html}}
}

@misc{cudaevents2024,
  title        = {CUDA Runtime API - Event Management},
  author       = {{NVIDIA Corporation}},
  year         = {2024},
  howpublished = {\url{https://docs.nvidia.com/cuda/cuda-runtime-api/group__CUDART__EVENT.html}}
}

@misc{NVIDIA_NCCL_User_Guide,
  author       = {{NVIDIA Corporation}},
  title        = {NVIDIA NCCL User Guide: Environment Variables},
  year         = 2025,
  howpublished = {\url{https://docs.nvidia.com/deeplearning/nccl/user-guide/docs/env.html#nccl-proto}},
  note         = {Accessed: 2025-01-11}
}

@misc{antgroup,
  title        = {Ant Group},
  howpublished = {\url{https://www.antgroup.com/en}},
  note         = {Accessed: 2025-01-13}
}

@misc{lfaidata,
  title        = {LF AI \& Data Foundation},
  howpublished = {\url{https://lfaidata.foundation/}},
  note         = {Accessed: 2025-01-13}
}

@misc{dlrover,
  author       = {Qinlong Wang and others},
  title        = {DLRover: An Automatic Distributed Deep Learning System},
  year         = 2023,
  howpublished = {\url{https://github.com/intelligent-machine-learning/dlrover}},
  note         = {Accessed: 2025-01-13}
}

@article{ramdas2017wasserstein,
  title={On wasserstein two-sample testing and related families of nonparametric tests},
  author={Ramdas, Aaditya and Garc{\'\i}a Trillos, Nicol{\'a}s and Cuturi, Marco},
  journal={Entropy},
  volume={19},
  number={2},
  pages={47},
  year={2017},
  publisher={MDPI}
}


%% file: references/papers.bib
@inproceedings{bergerTriangulatingPython,
  title = {Triangulating Python Performance Issues with {{SCALENE}}},
  booktitle = {17th {{USENIX Symposium}} on {{Operating Systems Design}} and {{Implementation}} ({{OSDI}} 23)},
  year = {2023},
  pages = {51--64},
  urldate = {2024-11-14},
  isbn = {978-1-939133-34-2},
  langid = {english},
  author = {Berger, Emery D. and Stern, Sam and Pizzorno, Juan Altmayer}
}

@article{daoFlashAttentionFast,
  title = {{{FlashAttention}}: Fast and Memory-Efficient Exact Attention with {{IO-awareness}}},
  year = {2022},
  month = jun,
  number = {arXiv:2205.14135},
  eprint = {2205.14135},
  primaryclass = {cs},
  publisher = {arXiv},
  urldate = {2024-04-29},
  archiveprefix = {arXiv},
  langid = {english},
  author = {Dao, Tri and Fu, Daniel Y. and Ermon, Stefano and Rudra, Atri and R{\'e}, Christopher}
}

@article{deepseek-aiDeepSeekV3Technical,
  title = {{{DeepSeek-V3}} Technical Report},
  year = {2024},
  month = dec,
  number = {arXiv:2412.19437},
  eprint = {2412.19437},
  primaryclass = {cs},
  publisher = {arXiv},
  doi = {10.48550/arXiv.2412.19437},
  urldate = {2025-01-09},
  archiveprefix = {arXiv},
  langid = {english},
  author = {{DeepSeek-AI} and Liu, Aixin and Feng, Bei and Xue, Bing and Wang, Bingxuan and Wu, Bochao and Lu, Chengda and Zhao, Chenggang and Deng, Chengqi and Zhang, Chenyu and Ruan, Chong and Dai, Damai and Guo, Daya and Yang, Dejian and Chen, Deli and {et al.}}
}

@inproceedings{dongEnhancingLargescale,
  title = {Enhancing Large-Scale {{AI}} Training Efficiency: The {{C4}} Solution for Real-Time Anomaly Detection and Communication Optimization},
  booktitle = {2025 {{IEEE International Symposium}} on {{High Performance Computer Architecture}} ({{HPCA}})},
  year = {2025},
  month = mar,
  pages = {1246--1258},
  issn = {2378-203X},
  doi = {10.1109/HPCA61900.2025.00095},
  urldate = {2025-07-11},
  langid = {english},
  author = {Dong, Jianbo and Luo, Bin and Zhang, Jun and Zhang, Pengcheng and Feng, Fei and Zhu, Yikai and Liu, Ang and Chen, Zian and Shi, Yi and Jiao, Hairong and Lu, Gang and Guan, Yu and Zhai, Ennan and Xiao, Wencong and Zhao, Hanyu and {et al.}}
}

@article{hanParameterEfficientFineTuning,
  title = {Parameter-{{Efficient Fine-Tuning}} for {{Large Models}}: {{A Comprehensive Survey}}},
  year = {2024},
  month = apr,
  number = {arXiv:2403.14608},
  eprint = {2403.14608},
  primaryclass = {cs},
  publisher = {arXiv},
  urldate = {2024-05-29},
  archiveprefix = {arXiv},
  langid = {american},
  author = {Han, Zeyu and Gao, Chao and Liu, Jinyang and Zhang, Jeff and Zhang, Sai Qian}
}

@inproceedings{ivchenkoTorchRecPyTorch,
  title = {{{TorchRec}}: A {{PyTorch}} Domain Library for Recommendation Systems},
  booktitle = {Proceedings of the 16th {{ACM Conference}} on {{Recommender Systems}}},
  year = {2022},
  month = sep,
  series = {{{RecSys}} '22},
  pages = {482--483},
  publisher = {Association for Computing Machinery},
  address = {New York, NY, USA},
  doi = {10.1145/3523227.3547387},
  urldate = {2025-01-09},
  isbn = {978-1-4503-9278-5},
  langid = {english},
  author = {Ivchenko, Dmytro and Van Der Staay, Dennis and Taylor, Colin and Liu, Xing and Feng, Will and Kindi, Rahul and Sudarshan, Anirudh and Sefati, Shahin}
}

@inproceedings{jiangMegaScaleScaling,
  title = {{{MegaScale}}: Scaling Large Language Model Training to More than 10,000 {{GPUs}}},
  booktitle = {21st {{USENIX Symposium}} on {{Networked Systems Design}} and {{Implementation}} ({{NSDI}} 24)},
  year = {2024},
  pages = {745--760},
  urldate = {2025-04-09},
  isbn = {978-1-939133-39-7},
  langid = {english},
  author = {Jiang, Ziheng and Lin, Haibin and Zhong, Yinmin and Huang, Qi and Chen, Yangrui and Zhang, Zhi and Peng, Yanghua and Li, Xiang and Xie, Cong and Nong, Shibiao and Jia, Yulu and He, Sun and Chen, Hongmin and Bai, Zhihao and Hou, Qi and {et al.}}
}

@article{lepikhinGshardScaling,
  title = {Gshard: Scaling Giant Models with Condi- Tional Computation and Automatic Sharding},
  year = {2021},
  langid = {english},
  author = {Lepikhin, Dmitry and Lee, HyoukJoong and Xu, Yuanzhong and Chen, Dehao and Firat, Orhan and Huang, Yanping and Krikun, Maxim and Shazeer, Noam and Chen, Zhifeng}
}

@article{liangTorchTitanOnestop,
  title = {{{TorchTitan}}: One-Stop {{PyTorch}} Native Solution for Production Ready {{LLM}} Pre-Training},
  year = {2024},
  month = nov,
  number = {arXiv:2410.06511},
  eprint = {2410.06511},
  primaryclass = {cs},
  publisher = {arXiv},
  doi = {10.48550/arXiv.2410.06511},
  urldate = {2025-01-03},
  archiveprefix = {arXiv},
  langid = {english},
  author = {Liang, Wanchao and Liu, Tianyu and Wright, Less and Constable, Will and Gu, Andrew and Huang, Chien-Chin and Zhang, Iris and Feng, Wei and Huang, Howard and Wang, Junjie and Purandare, Sanket and Nadathur, Gokul and Idreos, Stratos}
}

@inproceedings{narayananPipeDreamGeneralized,
  title = {{{PipeDream}}: Generalized Pipeline Parallelism for {{DNN}} Training},
  booktitle = {Proceedings of the 27th {{ACM Symposium}} on {{Operating Systems Principles}}},
  year = {2019},
  month = oct,
  pages = {1--15},
  publisher = {ACM},
  address = {Huntsville Ontario Canada},
  doi = {10.1145/3341301.3359646},
  urldate = {2024-04-29},
  isbn = {978-1-4503-6873-5},
  langid = {english},
  author = {Narayanan, Deepak and Harlap, Aaron and Phanishayee, Amar and Seshadri, Vivek and Devanur, Nikhil R. and Ganger, Gregory R. and Gibbons, Phillip B. and Zaharia, Matei}
}

@inproceedings{paszkePyTorchImperative,
  title = {{{PyTorch}}: An Imperative Style, High-Performance Deep Learning Library},
  booktitle = {Proceedings of the 33rd {{International Conference}} on {{Neural Information Processing Systems}}},
  year = {2019},
  month = dec,
  pages = {8026--8037},
  publisher = {Curran Associates Inc.},
  address = {Red Hook, NY, USA},
  urldate = {2025-01-01},
  langid = {english},
  author = {Paszke, Adam and Gross, Sam and Massa, Francisco and Lerer, Adam and Bradbury, James and Chanan, Gregory and Killeen, Trevor and Lin, Zeming and Gimelshein, Natalia and Antiga, Luca and Desmaison, Alban and K{\"o}pf, Andreas and Yang, Edward and DeVito, Zach and Raison, Martin and {et al.}}
}

@inproceedings{rajbhandariZeROMemory,
  title = {{{ZeRO}}: Memory Optimizations toward Training Trillion Parameter Models},
  booktitle = {{{SC20}}: {{International Conference}} for {{High Performance Computing}}, {{Networking}}, {{Storage}} and {{Analysis}}},
  year = {2020},
  month = nov,
  pages = {1--16},
  publisher = {IEEE},
  address = {Atlanta, GA, USA},
  doi = {10.1109/SC41405.2020.00024},
  urldate = {2024-04-29},
  copyright = {https://ieeexplore.ieee.org/Xplorehelp/downloads/license-information/IEEE.html},
  isbn = {978-1-7281-9998-6},
  langid = {english},
  author = {Rajbhandari, Samyam and Rasley, Jeff and Ruwase, Olatunji and He, Yuxiong}
}

@article{shengHybridFlowFlexible,
  title = {{{HybridFlow}}: A Flexible and Efficient {{RLHF}} Framework},
  year = {2024},
  month = oct,
  number = {arXiv:2409.19256},
  eprint = {2409.19256},
  publisher = {arXiv},
  urldate = {2024-11-18},
  archiveprefix = {arXiv},
  langid = {english},
  author = {Sheng, Guangming and Zhang, Chi and Ye, Zilingfeng and Wu, Xibin and Zhang, Wang and Zhang, Ru and Peng, Yanghua and Lin, Haibin and Wu, Chuan}
}

@article{shoeybiMegatronLMTraining,
  title = {Megatron-{{LM}}: Training Multi-Billion Parameter Language Models Using Model Parallelism},
  year = {2020},
  month = mar,
  number = {arXiv:1909.08053},
  eprint = {1909.08053},
  primaryclass = {cs},
  publisher = {arXiv},
  urldate = {2024-04-29},
  archiveprefix = {arXiv},
  langid = {english},
  author = {Shoeybi, Mohammad and Patwary, Mostofa and Puri, Raul and LeGresley, Patrick and Casper, Jared and Catanzaro, Bryan}
}

@inproceedings{wangDiagnosingApplicationnetwork,
  title = {Diagnosing Application-Network Anomalies for Millions of {{IPs}} in Production Clouds},
  booktitle = {2024 {{USENIX Annual Technical Conference}} ({{USENIX ATC}} 24)},
  year = {2024},
  pages = {885--899},
  urldate = {2024-11-17},
  isbn = {978-1-939133-41-0},
  langid = {english},
  author = {Wang, Zhe and Hu, Huanwu and Kong, Linghe and Kang, Xinlei and Xiang, Qiao and Li, Jingxuan and Lu, Yang and Song, Zhuo and Yang, Peihao and Wu, Jiejian and Yang, Yong and Ma, Tao and Liu, Zheng and Zeng, Xianlong and Cai, Dennis and {et al.}}
}

@inproceedings{wangZeroOverhead,
  title = {Zero Overhead Monitoring for Cloud-Native Infrastructure Using {{RDMA}}},
  booktitle = {2022 {{USENIX Annual Technical Conference}} ({{USENIX ATC}} 22)},
  year = {2022},
  pages = {639--654},
  urldate = {2024-11-17},
  langid = {english},
  author = {Wang, Zhe and Ma, Teng and Kong, Linghe and Wen, Zhenzao and Li, Jingxuan and Song, Zhuo and Lu, Yang and Chen, Guihai and Cao, Wei}
}

@inproceedings{wuGREYHOUNDHunting,
  title = {{{GREYHOUND}}: Hunting Fail-Slows in Hybrid-Parallel Training at Scale},
  booktitle = {2025 {{USENIX Annual Technical Conference}} ({{USENIX ATC}} 25)},
  year = {2025},
  pages = {731--747},
  urldate = {2025-07-16},
  isbn = {978-1-939133-48-9},
  langid = {english},
  author = {Wu, Tianyuan and Wang, Wei and Yu, Yinghao and Yang, Siran and Wu, Wenchao and Duan, Qinkai and Yang, Guodong and Wang, Jiamang and Qu, Lin and Zhang, Liping}
}

@inproceedings{yangAAsclepiusMonitoring,
  title = {{{AAsclepius}}: Monitoring, Diagnosing, and Detouring at the Internet Peering Edge},
  booktitle = {2023 {{USENIX Annual Technical Conference}} ({{USENIX ATC}} 23)},
  year = {2023},
  pages = {655--671},
  urldate = {2024-11-17},
  isbn = {978-1-939133-35-9},
  langid = {english},
  author = {Yang, Kaicheng and Li, Yuanpeng and Long, Sheng and Yang, Tong and Miao, Ruijie and Zhao, Yikai and Ji, Chaoyang and Mi, Penghui and Yang, Guodong and Xie, Qiong and Wang, Hao and Wang, Yinhua and Deng, Bo and Liao, Zhiqiang and Huang, Chengqiang and {et al.}}
}

@inproceedings{yaoHolmesLocalizing,
  title = {Holmes: Localizing Irregularities in {{LLM}} Training with Mega-Scale {{GPU}} Clusters},
  booktitle = {22nd {{USENIX Symposium}} on {{Networked Systems Design}} and {{Implementation}} ({{NSDI}} 25)},
  year = {2025},
  pages = {523--540},
  urldate = {2025-07-16},
  isbn = {978-1-939133-46-5},
  langid = {english},
  author = {Yao, Zhiyi and Hu, Pengbo and Miao, Congcong and Jia, Xuya and Liang, Zuning and Xu, Yuedong and He, Chunzhi and Lu, Hao and Chen, Mingzhuo and Li, Xiang and He, Zekun and Wang, Yachen and Zou, Xianneng and Jiang, Junchen}
}

@article{zhaoPyTorchFSDP,
  title = {{{PyTorch FSDP}}: Experiences on Scaling Fully Sharded Data Parallel},
  year = {2023},
  month = aug,
  journal = {Proc. VLDB Endow.},
  volume = {16},
  number = {12},
  pages = {3848--3860},
  issn = {2150-8097},
  doi = {10.14778/3611540.3611569},
  urldate = {2025-01-02},
  langid = {english},
  author = {Zhao, Yanli and Gu, Andrew and Varma, Rohan and Luo, Liang and Huang, Chien-Chin and Xu, Min and Wright, Less and Shojanazeri, Hamid and Ott, Myle and Shleifer, Sam and Desmaison, Alban and Balioglu, Can and Damania, Pritam and Nguyen, Bernard and Chauhan, Geeta and {et al.}}
}
